\documentclass[journal]{IEEEtran}
\usepackage[dvips]{graphicx}
\usepackage[cmex10]{amsmath}

\usepackage{mathrsfs}

\usepackage{comment}
\usepackage{bm}
\usepackage{array}
\usepackage{amsmath}
\usepackage{amscd}
\usepackage{amssymb}
\usepackage{latexsym}
\usepackage{epsfig}
\usepackage{cite}
\usepackage{setspace}
\usepackage{multirow}
\usepackage[english]{babel}
\usepackage[table]{xcolor}
\usepackage[margin=2.0cm]{geometry}
\usepackage{diagbox}

\usepackage{amsmath}
\usepackage{amscd}
\usepackage{amssymb}
\usepackage{latexsym}
\usepackage{epsfig}
\usepackage{cite}
\usepackage{setspace}
\usepackage{multirow}
\usepackage[english]{babel}
\usepackage[table]{xcolor}
\usepackage[margin=2.0cm]{geometry}

\newtheorem{ther}{Theorem}
\newtheorem{deft}{Definition}

\newtheorem{Proposition}{Proposition}
\newtheorem{Corollary}{Corollary}
\usepackage{mwe}

\ifCLASSINFOpdf

\else

\fi

\usepackage[tight,footnotesize]{subfigure}

\begin{document}
\title{Coprime Sensing via Chinese Remaindering over Quadratic Fields, Part II: Generalizations and Applications}
% Application of Chinese Remaindering over Rings: Two-Dimensional Direction-of-Arrival Estimation
\author{Conghui~Li,~
        Lu~Gan,~\IEEEmembership{Senior Member,~IEEE,}
        and~Cong~Ling,~\IEEEmembership{Member,~IEEE}% <-this % stops a space
\thanks{
%Manuscript received ; revised .  
This paper was presented in part at the tenth IEEE Sensor Array and Multichannel signal processing workshop (SAM), Sheffield, United Kingdom, July 8-11, 2018.}
\thanks{Conghui Li and Cong Ling are with the Department of Electrical and Electronic
Engineering, Imperial College London, London, SW7 2AZ, U.K. (e-mail:
conghui.li15@imperial.ac.uk; cling@ieee.org).}
\thanks{Lu Gan is with the Department of Electronic and Computer Engineering, Brunel University London, London, UB8 3PH, U.K. (e-mail: lu.gan@brunel.ac.uk).}
}
% change the date!!!!!!!!

\markboth{ }%
{Shell \MakeLowercase{\textit{et al.}}: 2-D DOA Estimation by Chinese Remaindering over Rings}

\maketitle

\begin{abstract}
The practical application of a new class of coprime arrays based on the Chinese remainder theorem (CRT) over quadratic fields is presented in this paper. 
The proposed CRT arrays are constructed by ideal lattices embedded from coprime quadratic integers with $\mathbf{B}_1 $ and $\mathbf{B}_2$ being their matrix representations respectively, whereby the degrees of freedom (DOF) surges to $O(|\det{(\mathbf{B}_1\mathbf{B}_2)}|)$ with $|\det (\mathbf{B}_1)| + |\det (\mathbf{B}_2)|$ sensors. The geometrical constructions and theoretical foundations were discussed in the accompanying paper in great detail, while this paper focuses on aspects of the application of the proposed arrays in two-dimensional (2D) remote sensing. 
A generalization of CRT arrays based on two or more pairwise coprime ideal lattices is proposed with closed-form expressions on sensor locations, the total number of sensors and the achievable DOF. The issues pertaining to the coprimality of any two quadratic integers are also addressed to explore all possible ideal lattices. Exploiting the symmetry of lattices, sensor reduction methods are discussed with the coarray remaining intact for economic maximization. 
In order to extend conventional angle estimation techniques based on uniformly distributed arrays to the method that can exploit any coarray configurations based on lattices, this paper introduces a hexagon-to-rectangular transformation to 2D spatial smoothing, providing the possibility of finding more compact sensor arrays. 
Examples are provided to verify the feasibility of the proposed methods. 
\nocite{li2018coprimeMIMO}

\end{abstract}


\IEEEpeerreviewmaketitle

\section{Introduction}

\IEEEPARstart{S} \normalfont parse arrays significantly increase the degrees of freedom by exploiting the concept of the coarray. As a result, they find a broad range of applications
% whereby a broad range of applications was found 
% in array signal processing 
including the direction of arrival estimation (DOA) \cite{gu2017optimized}, beamforming \cite{pal2010nested} and imaging \cite{bliss2003multiple}. A planar array consists of a set of antennas placed on a 2D lattice, which can estimate both azimuths and elevations of the impinging sources. 
In the accompanying paper \cite{li2018part1}, a new class of 2D sparse array configuration, namely the CRT array was introduced, which exploits quadratic integers to allocate antennas.    
Precisely, a CRT array is constructed by two coprime lattices generated from two quadratic integers respectively in the same field. According to the generalized CRT stated in \cite{Oggier10}, the difference vectors of two coprime lattices enjoy a quadratic surge of DOF, which allows the identification of more sources than the number of sensors.
The coprimality of lattices is guaranteed by the coprimality of their corresponding quadratic integers \cite[Theroem 1]{li2018part1}.
The theoretical foundations including the ring of quadratic integers and the issues pertaining to the prime factorization were also discussed in \cite{li2018part1}. 
One advantage of the CRT array is that it only requires a single prime integer $p$ for the prime factorization and offers simple closed-form representations of sensor locations, the number of sensors, the resulting DOF and so forth. Because of the limitations of conventional subspace-based algorithms, it is essential to enlarge the continuous coarray. A hole-free symmetric CRT array (HSCRT) with the enhanced coarray was proposed, which uses the notations of CRT arrays and modifies one of the subarrays by scaling its Voronoi cell \cite{li2018part1}.

In array processing, the basic idea of subspace-based algorithms is the orthogonality between the signal subspace and its corresponding noise subspace obtained from the eigendecomposition of the data covariance matrix \cite{schmidt1986multiple, roy1989esprit}. Therefore, by rearranging array manifold vectors and reprocessing signal outputs accordingly, the dimensions of the covariance matrices can be enhanced. Thus more sources can be detected from non-uniformly distributed array configurations including
\cite{chen2008minimum, pal2010nested,vaidyanathan2011sparse,vaidyanathan2011theory,pal2012nested,Qin15coprime,liu2016super,liu2017hourglass}. Among these, the sparse arrays including coprime arrays \cite{vaidyanathan2011sparse}, generalized coprime arrays \cite{Qin15coprime} and super nested arrays \cite{liu2016super} are more robust to mutual coupling compared to other one-dimensional (1D) array configurations such as minimum redundancy arrays \cite{chen2008minimum} and nested arrays \cite{pal2010nested}. While in the case of 2D sensor arrays, their applications include multidimensional DFT filter banks \cite{vaidyanathan2011theory}, angle estimations in the passive sensing scenario \cite{pal2012nested, liu2017hourglass} and so forth. Nevertheless, the application of 2D coprime arrays in the context of parameter estimations in MIMO radar has not been fully investigated.      

Because algebraic lattices are symmetric with respect to the origin, for any HSCRT, its sum coarray is identical to its difference coarray, which implies that HSCRT not only can be applied to active sensing (corresponding to the sum coarray) but also can be exploited for passive sensing (corresponding to the difference coarray). 
In this contribution, HSCRT is extended to new array configurations such as T, spinner, $\mathbb{Z}^2$ cross and $A_2$ cross to improve the essentialness \cite{liu2018SAM} while keeping the coarray intact as its corresponding HSCRT. Unlike HSCRT, such arrays are not symmetric with respect to the origin, but their subarrays have the symmetry with respect to the midpoint of two adjacent Voronoi cells.    

To verify the feasibility of the proposed arrays, we apply the CRT arrays to MIMO radar to estimate arrival angles whereby the half power beamwidth (HPBW) is significantly reduced, and the side lobes are largely suppressed.
On the other hand, in the passive sensing scenario, we discuss the applications of CRT arrays as multi-antenna receivers in the context of 2D DOA estimation with the presence of sensor mutual coupling. It has been proved in \cite{liu2016super} and \cite{liu2017hourglass} that the sparsity of the antennas is inversely related to the mutual coupling effect caused by electromagnetic interactions between adjacent antennas. As a result, one superiority of the proposed CRT arrays lies in the robustness to mutual coupling since all sensors placed on coprime lattices are well-separated.

The rest of the paper is outlined as follows. In Section \ref{A General Approach to Coprime Pairs of Quadratic Integers}, a general approach to coprime quadratic integers is proposed, providing a feasible method of generating coprime lattices.
Section \ref{Generalization of CRT array} generalizes CRT to any number of pairwise coprime algebraic integers.
By considering the symmetry of algebraic lattices, sliced CRT and cross arrays are introduced in Section \ref{Decrease the Number of Sensors} to improve the essentialness.
Based on the concept of 2D spatial smoothing, Section \ref{Applications to DOA Estimation} exploits the hexagon-to-rectangular technique for subspace-based algorithms. The superior performance of the proposed arrays is validated in Section \ref{Numerical Results}. Section \ref{Conclusion} concludes the paper.

\emph{Notations:} 
$R$ denotes a ring and $\Lambda$ is its algebraic lattice.
$i$ and $\omega$ denote the roots of $X^2+1=0$ and $X^2 - X + 1 =0$ respectively. $\mathbb{Z}[i]$ and $\mathbb{Z}[\omega]$ denote the rings of Gaussian and Eisenstein respectively. 
Bold font lowercase letters (e.g., $\mathbf{x}_1$), bold font uppercase letters (e.g., $\mathbf{G}$), fraktur font letters (e.g., $\mathfrak{p}_1$) and calligraphy font alphabets (e.g., $\mathcal{D}$) denote vectors, matrices, principal  ideals and sets respectively.
$\text{N}(m)=m\hat{m}$ denotes the norm of $m$ where $\hat{m}$ is the algebraic conjugate of $m$. $\mathcal{R}(m)$ and $\mathcal{J}(m)$ represent real and imaginary parts of $m$ respectively.

\section{A General Approach to Coprime Pairs of Quadratic Integers}\label{A General Approach to Coprime Pairs of Quadratic Integers}

To achieve a quadratic surge of DOF, the two lattices shall be coprime \cite{vaidyanathan2011theory,li2017coprime,li2018part1}.  
All of the CRT arrays introduced in the accompanying paper \cite{li2018part1} necessitate the use of pairwise coprime integers to construct coprime lattices. For this reason, the test of coprimality is essential in the coprime array design. A prime decomposition method and the coprime conditions on algebraic conjugate pairs were introduced in \cite{li2018part1}, by which two coprime lattices were generated. 

However, the number of ideals obtained from the prime decomposition is limited, and the existence of Bezout's identity is very difficult to verify when it comes to large complex numbers in general. In this section, we establish simplified necessary and sufficient conditions on relatively prime quadratic integers, which enrich the classes of coprime integers in number theory and offer more options on the selection of coprime lattices. 
Note that according to \cite[Theorem 1]{li2018part1}, the following theorem is equivalent to conditions on coprime matrices, which can be exploited not only in angle estimations but also in other applications such as beamforming and filter banks. 
The coprimality issues pertaining to Gaussian and Eisenstein integers are also addressed as special cases. 

\begin{ther}\label{coprime integers}
Let $\mathbb{Z}[q]$ denote a ring of integers with minimum polynomial $X^2+BX+C$. Two quadratic integers $m=m_1+m_2q$ and $n=n_1+n_2q$ in $\mathbb{Z}[q]$ are coprime if and only if
\begin{equation}\label{gcd(N(m), N(n), m_1n_1+m_2n_2)=1}
    \text{GCD}(\text{N}(m), \text{N}(n), m_1n_2-m_2n_1)=1 \,  \text{or equivalently}
\end{equation}
\begin{equation}\label{gcd(N(m), N(n), m_2n_1-m_1n_2)=1}
   \text{GCD}(\text{N}(m), \text{N}(n), m_1n_1- B m_1n_2+Cm_2n_2)=1,
\end{equation}
\end{ther}
where GCD denotes the greatest common divisor and $\text{N}(m)$ is the norm of $m$ that is defined by $\text{N}(m)=m\hat{m}=(m_1+m_2q)(m_1+m_2\hat{q})=m_1^2-m_1m_2B+m_2^2C$, $m_1,m_2 \in \mathbb{Z}$.

\emph{Proof}: 
See Appendix \ref{proof coprime integers}.
\hfill$\blacksquare$

Taking $\mathbb{Z}[\omega]$ and $\mathbb{Z}[i]$ as examples, the following two corollaries impose the coprime conditions on Eisenstein integers and Gaussian integers respectively.   
\begin{Corollary}
Two arbitrary Eisenstein integers $m$ and $n$ are relatively prime if and only if 
\begin{equation}\label{gcd11}
% gcd(m_1^2+m_1m_2+m_2^2, m_2n_1-m_1n_2,  n_1^2+n_1n_2+n_2^2) = 1.
%gcd(mm^*, m_2n_1-m_1n_2,  nn^*) = 1.
\text{GCD}(\text{N}(m), \text{N}(n), \frac{2}{\sqrt{3}}\operatorname{Im}(m\hat{n})) = 1,
\end{equation}
or equivalently,
\begin{equation}\label{gcd12}
\text{GCD}(\text{N}(m),\text{N}(n), \operatorname{Re}(m\hat{n}) - \frac{1}{\sqrt{3}}\operatorname{Im}(m\hat{n}) ) = 1.
\end{equation}
%$m=m_1+m_2\omega$ and $n=n_1+n_2\omega$
\emph{Proof}: Recall that Eisenstein integers are complex numbers of the form $m=m_1 + \omega m_2$ where $\omega$ and $\hat{\omega}$ are the roots of the polynomial $\omega^2-\omega+1=0$, i.e., $\omega=\frac{1}{2}+\frac{\sqrt{3}}{2}i$ and $\hat{\omega}=\frac{1}{2}-\frac{\sqrt{3}}{2}i$. 
By noticing that $m\hat{n}=(m_1n_1+\frac{1}{2}(m_1n_2+m_2n_1)+m_2n_2)+\frac{\sqrt{3}}{2}(m_2n_1-m_1n_2)i$, (\ref{gcd(N(m), N(n), m_1n_1+m_2n_2)=1}) can be simplified to (\ref{gcd11}). Similarly, substituting $B=-1$ and $C=1$ into (\ref{gcd(N(m), N(n), m_2n_1-m_1n_2)=1}) results in (\ref{gcd12}).      
\hfill$\blacksquare$
\end{Corollary}

\begin{Corollary}
Two arbitrary Gaussian integers $m$ and $n$ are relatively prime if and only if 
\begin{equation}\label{Ggcd11}
\text{GCD}(\text{N}(m), \text{N}(n), \operatorname{Im}(m\hat{n})) = 1,
\end{equation}
or equivalently,
\begin{equation}\label{Ggcd12}
\text{GCD}(\text{N}(m), \text{N}(n), \operatorname{Re}(m\hat{n})  ) = 1.
\end{equation}

\emph{Proof}: 
The minimum polynomial of Gaussian integers is $ X^2 +1 =0$ whereby all the integers are in the form of $m=m_1+m_2i$. Noticing that $m\hat{n}=(m_1n_1+m_2n_2)+(m_2n_1-m_1n_2)i$, the two coprimality conditions can be simplified to (\ref{Ggcd11}) and (\ref{Ggcd12}) respectively.   
\hfill$\blacksquare$
\end{Corollary}

In \cite[Theorem 1]{li2018part1}, it has been shown that if two quadratic integers in a principal ideal domain (PID) are coprime, their corresponding ideal lattices represented by matrices are relatively prime as well, i.e., the following two matrices: 
\begin{eqnarray*}
% \begin{aligned}
  & \mathbf{B}_m=\begin{pmatrix}
         m_1 &  -Cm_2\\
      m_2 &  m_1-Bm_2\\
  \end{pmatrix}
  \quad \text{and} \\
  &
  \mathbf{B}_{n}=\begin{pmatrix}
            n_1 &  -Cn_2\\
      n_2 &  n_1-Bn_2\\
  \end{pmatrix}
% \end{aligned}
\end{eqnarray*}
are coprime if and only if (\ref{gcd(N(m), N(n), m_1n_1+m_2n_2)=1}) or (\ref{gcd(N(m), N(n), m_2n_1-m_1n_2)=1}) holds.
Therefore, Theorem \ref{coprime integers} along with its corollaries further generalizes the coprimality conditions of algebraic lattices of dimension two to which CRT applies.

\section{$Q$-tuple CRT arrays}\label{Generalization of CRT array}

In \cite{li2018part1}, a novel method of designing sparse arrays was proposed by exploiting two prime ideals in quadratic number fields such as the fields of Gaussian and Eisenstein. Nevertheless, in general, the Chinese remaindering is applicable to any finite number of pairwise coprime integers which do not have to be prime elements and conjugate pairs. In this section, we will present the generalization of CRT arrays based on multiple quadratic integers. 

\begin{deft}
A $Q$-tuple CRT array consists of a number $Q$ of subarrays which are built from pairwise coprime ideals\footnote{Note that any two ideals $\mathcal{I}$ and $\mathcal{J}$ in a ring $R$ are coprime if and only if $\mathcal{I} + \mathcal{J} = R$, which implies that they are distinct \cite{li2018part1}.}
$\mathfrak{p}_k$ in a ring $R$ $(k=1 \cdots Q)$. Let $\mathfrak{P}=\prod^{Q}_{k=1}\mathfrak{p}_k$, with $\mathfrak{p}_k \neq \mathfrak{p}_j $ of $k \neq j$ being the factorization of the ideal $\mathfrak{P}$, then a $Q$-tuple CRT array incorporating with $Q$ ideal lattices can be characterized by 
\begin{equation*}
\mathcal{Z} =  \sigma (\mathfrak{p}_1) /  \sigma (\mathfrak{P}) \cup  \sigma (\mathfrak{p}_2) /  \sigma (\mathfrak{P}) \cdots \cup  \sigma (\mathfrak{p}_Q) /  \sigma (\mathfrak{P}) .   
\end{equation*}
\end{deft}
The virtual difference coarray of $\mathcal{Z}$ is an extension of the two-integer case defined in \cite[Definition 4]{pal2012nested1}, which can be represented by the set $\mathcal{D}$:
\begin{equation}\label{differencecoarraydef}
\mathcal{D} = \{ \mathbf{z}_m - \mathbf{z}_n \mid \mathbf{z}_m, \mathbf{z}_n \in \mathcal{Z} \}.
\end{equation}
The Chinese Remaindering Theorem for multiple ideals \cite[Theorem 3.5]{Oggier10} asserts that there exists a canonical isomorphism 
\begin{equation}\label{crtQ}
    R/\mathfrak{P} \simeq \prod^{Q}_{k=1} R/\mathfrak{p}_k,
\end{equation}
which implies that for all $ \{ a_1 \cdots a_Q \} \in\prod^{Q}_{k=1} R/\mathfrak{p}_k $, there exists $z \in R/\mathfrak{P}$ such that 
\begin{equation}\label{ZQK}
    z \equiv \sum^{Q}_{k=1}  a_k c_k \pmod{\mathfrak{P}}
\end{equation}
%Without loss of generality, the coprimality assumption on ideals   
where $ c_1 \cdots  c_Q \in R $ such that $c_k \equiv 1 \pmod{\mathfrak{p}_k}$ and $c_k \equiv 0 \pmod{\mathfrak{p}_j}$ for all $k \neq j $ as all ideals are pairwise coprime. Here $\{ c_1,  \cdots  c_Q \} $ serves as our CRT basis. Based on this, the next proposition derives the properties of the generalized $Q$-tuple CRT arrays.
%\cite{lyubashevsky2010ideal}

\begin{Proposition}
The number of sensors in the $Q$-tuple CRT array is given by 
\begin{equation}\label{ZQ1}
\begin{aligned} 
& |\mathcal{Z}| = \sum_{k_1=1}^{Q} \text{N} (\mathfrak{p}_{k_1}^{-1}\mathfrak{P}) - \sum_{ 1\leq k_1 < k_2 \leq Q} \text{N}((\mathfrak{p}_{k_1}\mathfrak{p}_{k_2})^{-1}\mathfrak{P}) \\
   &  + \sum_{1\leq k_1 < k_2 < k_3 \leq Q}\text{N}((\mathfrak{p}_{k_1}\mathfrak{p}_{k_2}\mathfrak{p}_{k_3})^{-1}\mathfrak{P})  \cdots  + (-1)^{Q-1} \\
 & = \sum_{k_1=1}^Q \frac{ | \det (\mathbf{B}_{\mathfrak{P}}) | }{| \det(\mathbf{B}_{k_1}) |} -  \sum_{1\leq k_1 < k_2 \leq Q } \frac{ | \det (\mathbf{B}_{\mathfrak{P}}) |}{ | \det(\mathbf{B}_{k_1}) \det(\mathbf{B}_{k_2} ) | } \cdots  \\
 & = \sum_{k=1}^Q (-1)^{k+1} \sum_{1\leq k_1 < \cdots < k_k \leq Q} \frac{ | \det (\mathbf{B}_{\mathfrak{P}}) |}{ 
 \prod_{j=1}^k |\det(\mathbf{B}_{k_j})| },  
\end{aligned}
\end{equation}
and the maximum DOF is
\begin{equation}
 | \mathcal{D} | =  \text{N} ( \mathfrak{P} )  = | \det (\mathbf{B}_{\mathfrak{P}}) |, 
\end{equation}
\end{Proposition}
where $ | \mathcal{D}|$, $\text{N}(\mathfrak{p}_k)$ and $| \det (\mathbf{B}_k) |$ denote the cardinality of $\mathcal{D}$, the norm of $\mathfrak{p}_k$ and the absolute value of the determinent of $\mathbf{B}_k$ respectively. $\mathbf{B}_{\mathfrak{P}}$ is the matrix representation of the ideal $\mathfrak{P}$ and $\mathbf{B}_k$ is the matrix representation of $\mathfrak{p}_k$ for all $1 \leq k \leq Q$.

\emph{Proof}: The total number of physical sensors parameterized by the norm of $\mathfrak{p}_k$ is the sum of all the identical lattice points within $\sigma (\mathfrak{p}_1) /  \sigma (\mathfrak{P}), \cdots \sigma (\mathfrak{p}_{Q-1}) /  \sigma (\mathfrak{P})$ and $ \sigma (\mathfrak{p}_Q) /  \sigma (\mathfrak{P})$. From lattice theory point of view \cite{conway2013sphere}, $ \sigma (\mathfrak{P})$ can be viewed as a sublattice of $\sigma (\mathfrak{p}_k)$. By \cite[Definition 3.12]{oggier2004algebraic}, the cardinality of the quotient group $\sigma (\mathfrak{p}_k) /  \sigma (\mathfrak{P})$ is the number of lattice points within this group and can be calculated as:
\begin{equation*}
    | \sigma (\mathfrak{p}_k) /  \sigma (\mathfrak{P}) | = \frac{|\det(\mathbf{B}_{\mathfrak{P}})|}{|\det(\mathbf{B}_k)|},
\end{equation*}
which is also the norm of $\mathfrak{p}_k^{-1}  \mathfrak{P}$ for all $1 \leq k \leq Q$ because the embedding $\sigma$ is bijective and $\text{N} (\mathfrak{p}_k^{-1} \mathfrak{P}  ) = \text{ N} ( \mathfrak{P} ) /\text{N} (\mathfrak{p}_k )  $ \cite[Corollary 2.12]{Oggier102}.

Due to the assumption that all the ideals are pairwise coprime, i.e., $ \mathfrak{p}_k + \mathfrak{p}_j = R $ for all $k \neq j $, there are no duplicate elements excluding the origin $\mathbf{0}$ and thier products.
Therefore, the calculation is executed by the inclusion-exclusion principle \cite{o1993inclusion}.
For example, with $Q=3$ there are a number $\text{N}((\mathfrak{p}_1 \mathfrak{p}_2)^{-1} \mathfrak{P})$ of the multiplications $\text{N}(\mathfrak{p}_1\mathfrak{p}_2)$ in $\text{N}(\mathfrak{P})$ and the process ends when $ \text{N}((\mathfrak{p}_1\mathfrak{p}_2\mathfrak{p}_3)^{-1}\mathfrak{P}) = 1 $.  
With $Q=2$, there are no duplicate sensors except $\mathbf{0}$ since $\mathfrak{p}_1$ and $\mathfrak{p}_2$ are relatively prime. (\ref{ZQ1}) becomes 
$ \text{N}(\mathfrak{p}_1) + \text{N}(\mathfrak{p}_1) - \text{N}((\mathfrak{p}_1\mathfrak{p}_2)^{-1}\mathfrak{P}) = \text{N}(\mathfrak{p}_1) + \text{N}(\mathfrak{p}_1) -1  $, which coincides with \cite[Proposition 1]{li2018part1} with $\mathfrak{p}_1= \langle \, m \, \rangle $, $\mathfrak{p}_2 = \langle \, n \, \rangle$ and $\mathfrak{P} = \langle \, m n \, \rangle = \langle \, p \, \rangle$.  

According to the ring isomorphism (\ref{crtQ}), the solutions of $z$ are unique, which can also be verified by extending the case of two coprime ideals given in \cite[Proposition 1]{li2018part1} by setting $\mathcal{I}=\mathfrak{p}_k$ and $\mathcal{J}=(\mathfrak{p}_k)^{-1}\mathfrak{P} $.  
 
\hfill$\blacksquare$

%  \begin{figure}[tb]
%       \centering
%      %\framebox{\parbox{3in}{content}}
%      \includegraphics[height=1.8in]{Figs/Jx1selection2.png}
%     %\includegraphics[scale=0.4]{Figs/Jx1selection.png}
%     % \vspace{-2mm}
%       \caption{A graphical representation of elements in $\mathcal{S}_{0,0}$ with $l_R=7$ and $l_p=3$ from which $\mathcal{J}_{x1}$ can be selected by matrix $\mathbf{J}_{x1}$. Physical positions of sensors on the edges are shown by coordinates in the $\mathbf{u}$ space.}
%       \label{Jx1selection}
%       \vspace{-3mm}
%   \end{figure}

 \begin{figure}[tb]
    \centering
    \includegraphics[scale=0.58]{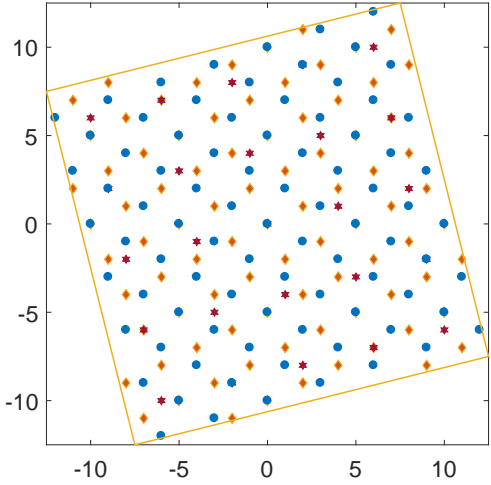}
    \caption{An example of a multi-lattice array constructed from three pairwise coprime algebraic integers. $\mathfrak{p}_1= \langle \, -1-2i \, \rangle  $ (orange diamonds), $\mathfrak{p}_2=\langle \,-1+2i \, \rangle$ (blue dots), $\mathfrak{p}_3=\langle \, -1+4i \, \rangle $ (red hexagrams) and $\mathfrak{P}$ as the Voronoi cell (yellow polygon).}
    \label{threeinteger}
% \vspace{-6mm}
\end{figure}

Considering a PID, for example, in $\mathbb{Z}$, 2,  3 and 7 are pairwise coprime, thus $\langle \, 2 \, \rangle = 2 \mathbb{Z}$,  $\langle \, 3 \, \rangle  = 3 \mathbb{Z} $ and $\langle \, 7 \, \rangle = 7 \mathbb{Z} $ are coprime ideals. According to CRT,  $\mathbb{Z}$ can be generated by a linear combination of these three ideals with coefficients in $\mathbb{Z}$, i.e., $ -\langle \, 2 \, \rangle - \langle \, 3 \, \rangle + \langle \, 7 \, \rangle = \mathbb{Z}$ where $1, -1 \in \mathbb{Z}$. 
These three ideals form a $3$-tuple CRT array in 1D with $\mathfrak{p}_1 = \langle \, 2 \, \rangle$, $\mathfrak{p}_2 = \langle \, 3 \, \rangle $, $\mathfrak{p}_3 = \langle \, 7 \, \rangle $ and $\mathfrak{P}=\langle \, 42 \, \rangle$.
The total number of antennas is $\frac{42}{2} + \frac{42}{3} + \frac{42}{7} - (2+3+7) +1 = 30$ if only the left boundary is defined in this Voronoi cell. 
Analogously, in 2D, a 3-tuple CRT array consists of three pairwise coprime ideals in quadratic fields. Taking $\mathbb{Z}[i]$ as an example, by Theorem \ref{coprime integers}, $\langle \,-1+2i \, \rangle $,  $\langle \, -1+4i \, \rangle $ and  $\langle \, -1-2i \, \rangle $ are three pairwise coprime ideals whose matrix representations are \cite{li2018part1}:
\begin{equation*}
 \begin{aligned}
  & \mathbf{B}_1=\begin{pmatrix}
    -1 & -2\\  2  &   -1
  \end{pmatrix}
  , \quad
  \mathbf{B}_2=\begin{pmatrix}
    -1  &  -4  \\  4&-1
  \end{pmatrix}, \\
 &  \quad  \text{and} \quad
    \mathbf{B}_3=\begin{pmatrix}
    -1&2\\        -2&-1
  \end{pmatrix}.
\end{aligned}
\end{equation*}
In this case $\mathfrak{P}= \langle \,-1+2i \, \rangle \langle \, -1+4i \, \rangle \langle \, -1-2i \, \rangle = \langle \, -5+20i \, \rangle $ and $|\det(\mathbf{B}_{\mathfrak{P}}|) = \text{N}(\mathfrak{P})=425$. Thus the total number of sensors is 
\begin{equation*}
\begin{aligned}
 |\mathcal{Z}| & = \frac{|\det(\mathbf{B}_{\mathfrak{P}})| }{|\det(\mathbf{B}_{1}) |} +\frac{|\det(\mathbf{B}_{\mathfrak{P}})| }{|\det(\mathbf{B}_{2})| } +\frac{|\det(\mathbf{B}_{\mathfrak{P}})| }{|\det(\mathbf{B}_{3})| } \\ 
 & - \bigg( \frac{|\det(\mathbf{B}_{\mathfrak{P}})| }{|\det(\mathbf{B}_{1}\mathbf{B}_{2})| } +\frac{|\det(\mathbf{B}_{\mathfrak{P}})| }{|\det(\mathbf{B}_{2}\mathbf{B}_{3})| } +\frac{|\det(\mathbf{B}_{\mathfrak{P}})| }{|\det(\mathbf{B}_{1}\mathbf{B}_{3})| }\bigg) + 1\\
 & =\text{N}( 9-2i )  + \text{N}(5) + \text{N}(-7-6i)  \\
 & - \Big(\text{N}( -1+2i )+\text{N}( -1+4i )+\text{N}( -1-2i ) \Big)  +  1 \\  
 & = \frac{425}{5}+\frac{425}{17}+\frac{425}{5}-(17+5+5)+1=169.       
 \end{aligned}
\end{equation*}
Fig. \ref{threeinteger} depicts the array configuration of this 3-tuple CRT example.

\section{Economic Maximization of CRT Arrays} \label{Decrease the Number of Sensors}

A sensor in the array is called \emph{essential}, if the coarray configuration of this array varies between with and without it \cite{liu2017maximally}. Depending on the number of essential sensors in the array, the criterion of \emph{maximally economic arrays} was first proposed in \cite{liu2017maximally} and further discussed in \cite{liu2018SAM} along with trade-offs between the robustness and the size of the coarray.    
Due to the symmetry of ideal lattices, the number of sensors in HSCRT can be reduced without degrading the coarrays to boost the economy of the sensor arrays. 
In this section, we will improve essentialness of the CRT arrays that are based on $\mathbb{Z}^2$ lattice and $A_2$ lattice which can be inherited to other lattices. 

\subsection{Sliced CRT arrays}

Let us first recall the definition of the difference coarray: 
\begin{equation*}
  \mathcal{D} =   \{ \mathbf{d}_{m,n} = \mathbf{z}_m-\mathbf{z}_n \; | \; \mathbf{z}_n,\mathbf{z}_m \in \mathcal{Z}\} . 
\end{equation*}
Note that if the array configuration is non-symmetric with respect to the Voronoi central point, the difference vector $ \mathbf{z}_m - \mathbf{z}_n $ is different from  $ \mathbf{z}_n - \mathbf{z}_m $. 

The first economic maximization method is simply a reduction of the subarray with the larger aperture. 
Recall that HSCRT consists of two subarrays, namely $\sigma(\mathfrak{p}_1)\cap \mathcal{V}(pR)$ and $\sigma(\mathfrak{p}_2)\cap \mathcal{V}(2pR)$ where $p$ is a prime \cite{li2018part1}. A sliced CRT array is a modified version of HSCRT with the number of elements in one of the subarrays reduced to half. Precisely, we partition the larger Voronoi cell $\sigma(\mathfrak{p}_2)\cap \mathcal{V}(2pR)$ into $r$ sectors ($r$ is even) with equal area so that these sectors are symmetric with respect to its Voronoi central in pairs. Then it is feasible to remove the points in one sector of each pair without degrading the performance, i.e., the consecutive array aperture remains intact. Considering underlying symmetries of $\mathbb{Z}^2$ and $A_2$, we select $r=2$ to all $\mathbb{Z}^2$-based arrays and $r=6$ to all $A_2$-based arrays, after which their array configurations resemble T and spinner respectively. Let $ \mathcal{Z}_1= \sigma (\mathfrak{p}_1) \cap  \mathcal{V} \big(\sigma (\mathfrak{p}_1\mathfrak{p}_2 )\big)$ and $\mathcal{Z}_2= \sigma (\mathfrak{p}_2) \cap  \mathcal{V} \big( \sigma (\mathfrak{p}_1\mathfrak{p}_2 ) \big)$, the definitions of some examples of sliced arrays are given using Voronoi cells $\mathcal{V}$ as follows.

\begin{deft}[T arrays]
The T array configuration is given as: 
\begin{equation*}
    \mathcal{T} =  \mathcal{Z}_1 \cup \{ \mathbf{z}=(z_x, z_y) \, | \,  z_y > 0,  \mathbf{z} \in \mathcal{Z}_2 \}. 
\end{equation*}
\end{deft}

\begin{deft}[Spinner arrays]
The sensors of the spinner array are allocated on   
\begin{eqnarray*}
       & \mathcal{P} = \mathcal{Z}_1 \cup \{ \mathbf{z}=(z_x, z_y) \, | \,  0 < z_y < \sqrt{3}z_x, \\
       & \text{ or }  0 < z_y < -\sqrt{3}z_x, \text{ or }  z_y < \pm \sqrt{3}z_x ;  \mathbf{z} \in \mathcal{Z}_2   \}.
\end{eqnarray*}
\end{deft}

\begin{Proposition}\label{sliced CRT array}
Let $\Lambda$ denote the algebraic lattice of $R=\mathbb{Z}[q]$. The contiguous coarrays of sliced arrays include all lattice points in $\Lambda \cap \mathcal{V}(p\Lambda)$.  
\end{Proposition}

\emph{Proof}:
Let $\mathbf{z}_m$ and $\mathbf{z}_n$ be two arbitrary points in $ \mathcal{Z}_1 $ and  $ \mathcal{Z}_2$ respectively. The difference vector can be rewritten as:
\begin{equation*}
\mathbf{z}_m -\mathbf{z}_n =(-\mathbf{z}_n)-(-\mathbf{z}_m).
\end{equation*}
Because $\mathcal{Z}_1$ and $\mathcal{Z}_1$ are centrosymmetric, it can be verified that $-\mathbf{z}_m \in \mathcal{Z}_1$ and $-\mathbf{z}_n \in \mathcal{Z}_2$. Half of the elements in $ \mathcal{Z}_2$ become redundant as they do not contribute to the difference coarray and can be replaced by $(-\mathbf{z}_m)$. This implies if we remove half of the elements in $ \mathcal{Z}_2 $, the difference coarray is intact. For instance with $r=2$, the new subarray is 
$\mathcal{Z}'_2 = \{ \mathbf{z} =(z_x, z_y) \; | \; \mathbf{z}  \in \mathcal{Z}_2, z_y > 0 \}$ while $\mathcal{Z}_1$ remains the same. 
Elements in the set $\mathcal{Z}''_2 = \mathcal{Z}_2-\mathcal{Z}'_2= \{ \mathbf{z} \in  \mathcal{Z}_2 \, \text{and} \, \mathbf{z} \notin \mathcal{Z}'_2 \} $ are redundant regarding the contribution to the difference coarray because 
$ \mathbf{z}_j{''}  -  \mathbf{z}_n$ can be replaced  by $(-\mathbf{z}_n) - \mathbf{z}'_m$ for all $ \mathbf{z}_j{''} \in \mathcal{Z}''_2  $ and $\mathbf{z}_m{'} \in \mathcal{Z}'_2 $.
\hfill$\blacksquare$
 
% The proofs for other values of $r$ is similar to $r=2$, the only difference is the configuration of $ \mathcal{Z}'_1  $. 
The properties of sliced CRT arrays can be derived as follows: Assuming the two subarrays are generated by two coprime ideals $\mathfrak{p}_1$ and $\mathfrak{p}_2$, the total number of sensors is $ 2 (\text{N}(\mathfrak{p}_1)-1) + \text{N}(\mathfrak{p}_2)$ and also equals $3p-2$ if the two ideals are obtained from the decomposition of prime $p$ \cite{li2018part1}. Thus the total decrease of sensor number is $2(\text{N}(\mathfrak{p}_1)-1)$ compared to HSCRT.

\begin{figure}[tb]
  \begin{center}
    \subfigure[]{\includegraphics[height=2.2in]{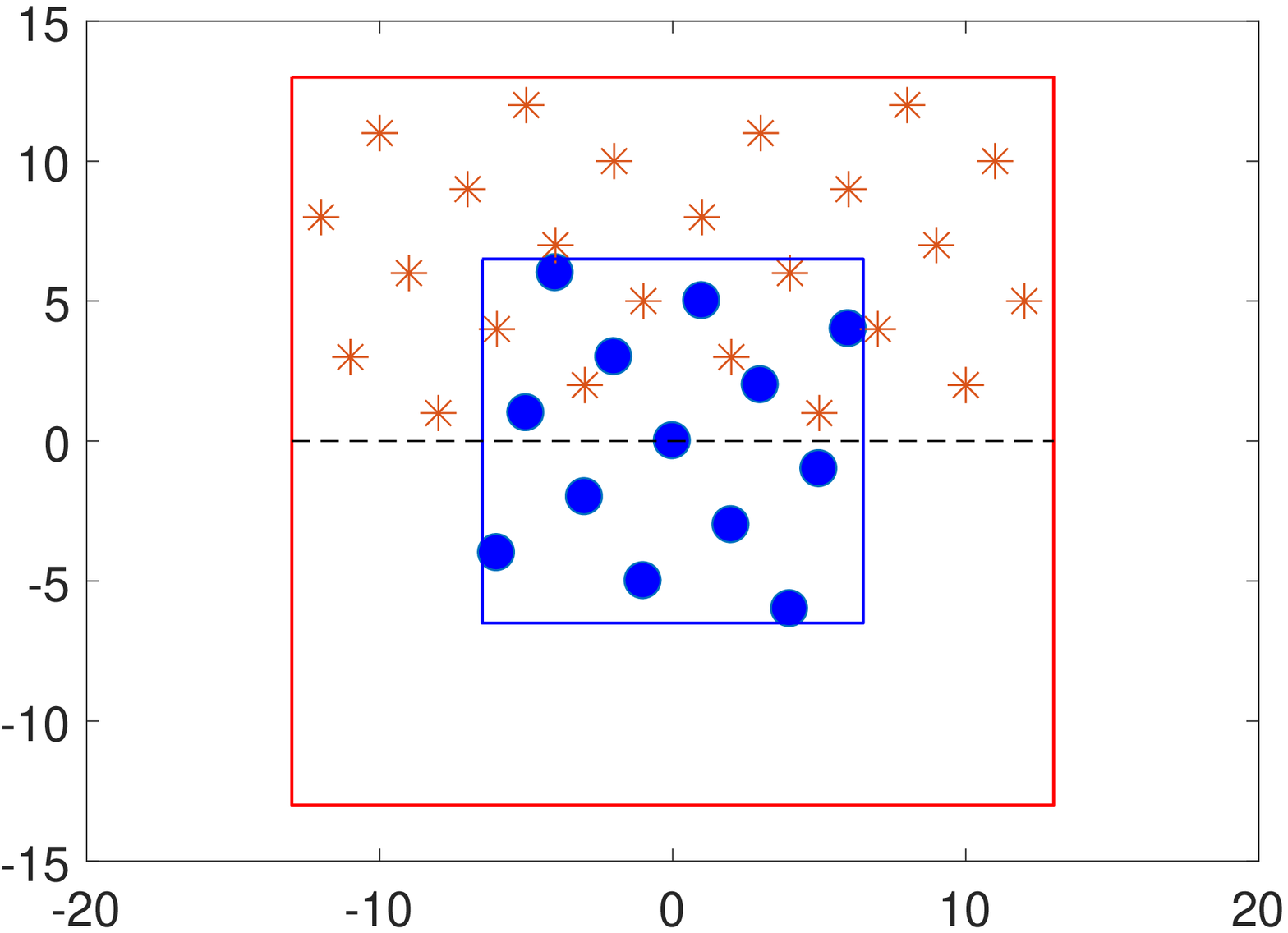}\label{g13dpositions1}}
     \subfigure[]{\includegraphics[height=2.2in]{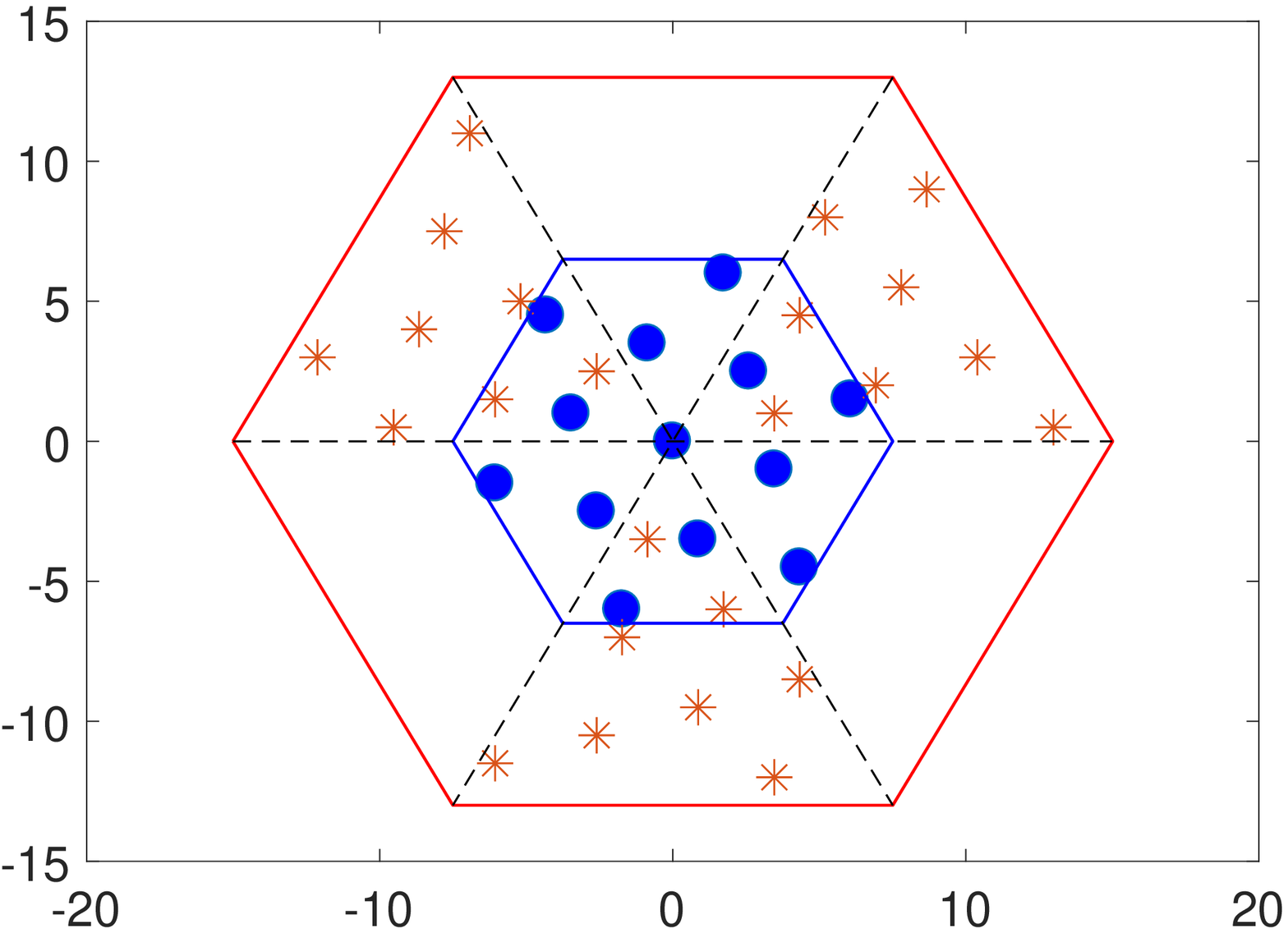}\label{e13dpositions}}
     \vspace{-1mm}
    \caption{The T array configuration from the decomposition of $p=13$ (a) over Gaussian integers with sensors in $\sigma(\langle \, 3+2i \, \rangle)/13\mathbb{Z}^2$ shown in blue dots and reduced sensors in $ \sigma( \langle \, 3-2i \, \rangle)/26\mathbb{Z}^2$ in red stars. The spinner shape configuration of $p=13$ (b) over Eisenstein integers with $\sigma(\langle \, 1+2\sqrt{3}i \, \rangle)/13A_2$ in blue dots and reduced sensors in $\sigma( \langle \,  1-2\sqrt{3}i \, \rangle)/26A_2$ in red stars. The Voronoi cells $\mathcal{V}(13\Lambda)$ and $\mathcal{V}(26\Lambda)$ are shown in red and blue.} \label{g5reduce}
    \label{sector}
  \end{center}
\vspace{-2em}
\end{figure}

Examples of T array and spinner array are illustrated in Fig. \ref{sector} where the symmetry is indicated by black dashed lines.

 \begin{figure}[tb]
  \begin{center}
      \subfigure[]{\includegraphics[height=2.2in]{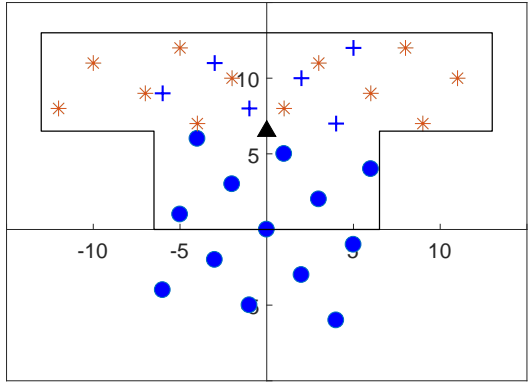}\label{cross22}}
      \subfigure[]{\includegraphics[height=2.2in]{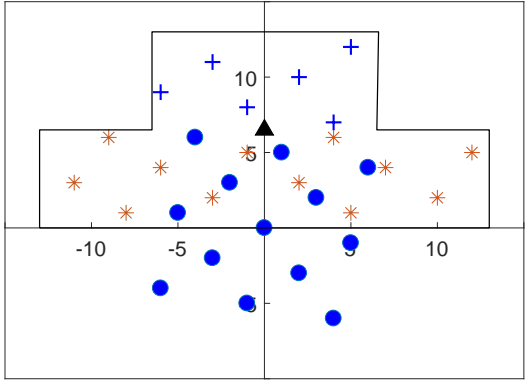}\label{cross33}} 
    \vspace{-1mm}
    \caption{Illustration of the symmetry for $\mathbb{Z}^2$ array. The extend points of $\sigma(\mathfrak{p}_2)/13\mathbb{Z}^2$ are shown in '$+$' which are symmetrical with respect to the midpoint of the Voronoi edge. The midpoint is shown in the small triangle. The remaining half of $ 2\sigma(\mathfrak{p}_1)/13\mathbb{Z}^2$ in red which are also symmetric with respect to the triangle point.}
    \label{crossd}
  \end{center}
\vspace{-6mm}
\end{figure}

 \begin{figure}[tb]
  \begin{center}
    \subfigure[]{\includegraphics[height=2.2in]{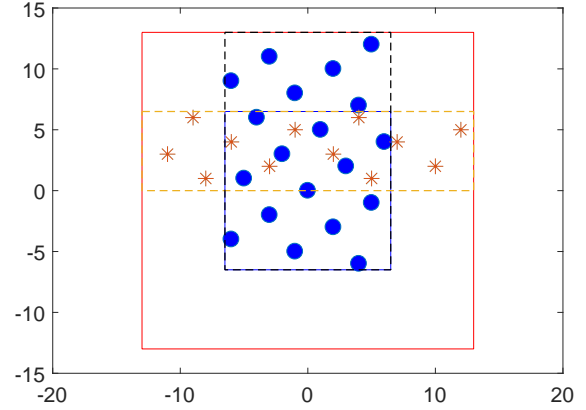}\label{crossg13}}
     \subfigure[]{\includegraphics[height=2.2in]{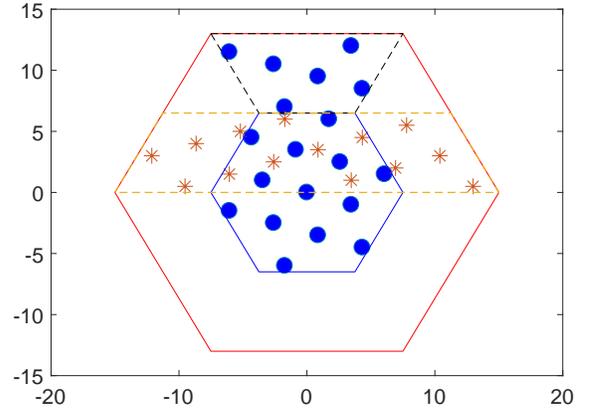}\label{crosse13}}
     \vspace{-1mm}
    \caption{$\mathbb{Z}^2$ cross array (a) and $A_2$ cross array (b) with $p=13$. The Voronoi cells $\mathcal{V}(13\Lambda)$ and $\mathcal{V}(26\Lambda)$ are shown in red and blue with the symmetric extension shown as black dashed lines and remaining part of $\mathcal{V}(26\Lambda)$ as yellow dashed lines.} 
    \label{cross}
  \end{center}
\vspace{-1em}
\end{figure}

\subsection{Cross Arrays}
In this subsection, we introduce cross shape array configurations with further sensor reduction by exploiting the underlying symmetry of the Voronoi cells. 
According to Proposition \ref{sliced CRT array}, all sensors below the $x$ axis in the larger array can be removed, after which the remaining points in the larger Voronoi cell $\sigma(\mathfrak{p}_2) \cap \mathcal{V}(2p\Lambda)$ have the symmetry with respect to the corresponding centroid of two adjacent Voronoi central points. Note that the center of the mass is also the generating point of each cell. Therefore this center of symmetry is identical to the midpoints of edge segments of the Voronoi cell $\mathcal{V}(p\Lambda)$. For example, the points in the set 
$\{ \mathbf{z}=(z_x, z_y) \; | \; \mathbf{z} \in \sigma(\mathfrak{p}_2)/26\mathbb{Z}^2,  \quad z_y \geq 0 \}$ 
are symmetric with respect to the midpoint of the edge of $\mathcal{V}(13\mathbb{Z}^2)$, namely $(0, \frac{13}{2})$. Regarding the midpoint as the center of symmetry, the redundant sensors in the larger array can be removed after extending the smaller array symmetrically with respect to the midpoint of the edge segment as compensation to the coarray, thus it comes to the term of the cross array. For instance, it can be verified that after the extension of the smaller cell, the points inside the $\mathbf{\top}$ shape as shown in Fig. \ref{cross22} and the points inside the $ \mathbf{\bot } $ shape as shown in Fig. \ref{cross33} yield the same difference coarray in the small Voronoi cell.
In the interest of the intact virtual coarray, 
Fig. \ref{cross33} is adequate for our purpose.

Let us denote $(x_x,x_y)$ as the position of the centroid of two adjacent Voronoi central points of $\mathcal{V}(p\Lambda)$ and
$\mathcal{H} = \{ \mathbf{h}=(h_x, h_y) \; | \; \mathbf{h} \in \mathcal{Z}_1, h_y > 0 \}$ for simplicity.

\begin{deft}[Cross arrays]
The sensors positions of the cross array are explicitly given by $(z_x,z_y) \in \mathcal{C}$ where
\begin{eqnarray*}
       & \mathcal{C} =\mathcal{Z}_1 \cup \{ \mathbf{z}= (z_x, z_y) \; | \;  0 < z_y < x_x , \mathbf{z} \in \mathcal{Z}_2 \} \cup \\ 
       & \{  (z_x, z_y)  \, | \, z_x=2x_x-h_x, z_y=2x_y-h_y, (h_x, h_y)\in \mathcal{H}\} \\
\end{eqnarray*}
\end{deft}

\begin{Proposition}
The cross array can generate all lattice points in $\Lambda \cap \mathcal{V}(p\Lambda)$.  
\end{Proposition}

\emph{Proof}:
Let $\mathbf{c}  \in \mathcal{C}$ and $\mathbf{e} \in \mathcal{E}$ where 
$\mathcal{C} = \{ \mathbf{c}=(c_x,c_y) \; |\; \mathbf{c} \in \Lambda \cap \mathcal{V}(2\sigma(\mathfrak{p}_1)), c_x \geq 0  \}$, and $ \mathcal{E} = \Lambda \cap \mathcal{V}(\sigma(\mathfrak{p}_2))$. 
 Denote $\mathbf{G}_1$ and $\mathbf{G}_2$ as generator matrices of $\sigma(\mathfrak{p}_1)$ and $\sigma(\mathfrak{p}_2)$ respectively. 
According to the commutativity of matrix representations \cite[Lemma 2]{li2018part1} ($\mathbf{B}_1\mathbf{B}_2=\mathbf{B}_2\mathbf{B}_1$), an arbitrary element in the difference coarray can be expressed as  
\begin{equation}\label{zalpha}
\begin{aligned} 
      & \mathbf{d}  = \mathbf{G}_2\mathbf{c} - \mathbf{G}_1\mathbf{e} = \mathbf{G}( \mathbf{B}_2\mathbf{c} - \mathbf{B}_1\mathbf{e}) \\ 
      & = \mathbf{G} \big( \mathbf{B}_1( \mathbf{B}_2\mathbf{b} - \mathbf{e} ) - \mathbf{B}_2( \mathbf{B}_1\mathbf{b} - \mathbf{c}) \big) \\
      & = \mathbf{G} \big( \mathbf{B}_1\mathbf{e}' - \mathbf{B}_2\mathbf{c}' \big)=  \mathbf{G}_1\mathbf{e}' - \mathbf{G}_2\mathbf{c}' ,
\end{aligned}
\end{equation}
where $\mathbf{b}\triangleq 2\mathbf{B}_1^{-1}\mathbf{B}_2^{-1}\mathbf{x}$ and $\mathbf{G}$ is the generator matrix of $\Lambda$ \cite[Section II-B]{li2018part1}. By varying the vector $\mathbf{c}$ such that it exhausts all values in $\mathcal{C}$ respectively, $\mathbf{c}'\triangleq \mathbf{B}_1\mathbf{b} - \mathbf{c}$ varies accordingly.
Left multiplying $\mathbf{G}\mathbf{B}_2$ to $\mathbf{c}'$ yields $\mathbf{G}\mathbf{B}_2\mathbf{c}'=\mathbf{G}\mathbf{B}_1\mathbf{B}_2\mathbf{b}-\mathbf{G}\mathbf{B}_2\mathbf{c}$, i.e.
\begin{equation*}
   \frac{1}{2}(\mathbf{G}_2\mathbf{c} + \mathbf{G}_2\mathbf{c}')= \mathbf{G}\mathbf{x},
\end{equation*}
which implies that for all $\mathbf{c} \in \mathcal{C}$, $\mathbf{G}_2\mathbf{c}$ is symmetric to $\mathbf{G}_2\mathbf{c}'$ with respect to $\mathbf{G}\mathbf{x}$. 
Induced by the symmetry of the Voronoi cell, $\mathbf{G}_2\mathbf{c}'$ is in $\mathcal{C}$ for all $\mathbf{c}'=\mathbf{B}_1\mathbf{b} - \mathbf{c}$ and can be removed. 

Likewise, $ \mathbf{G}_1\mathbf{e}'$ and $\mathbf{G}_1\mathbf{e}$ are symmetric with respect to $\mathbf{G}\mathbf{x}$ ($\mathbf{G}_1\mathbf{e} + \mathbf{G}_1\mathbf{e}'= 2\mathbf{G}\mathbf{x}$) and by exhausting $a \in \mathcal{E}$, $\mathbf{G}_1\mathbf{e}' = 2\mathbf{G}\mathbf{x}^T - \mathbf{G}_1\mathbf{e} $ gives the positions of extending points in the cross array.  
\hfill$\blacksquare$

In the example for the $\mathbb{Z}^2$ cross array shown in Fig. \ref{crossd}, the sensor reduction can be executed by removing half of the points in the set $\Lambda \cap \mathcal{V}(2\sigma(\mathfrak{p}_1))$ to get $\mathcal{C}$ and introducing $\mathbf{e}'=2\mathbf{G}_1^{-1}\mathbf{x}^T-\mathbf{e}$ into $\mathcal{E}$, whereby the total number of physical sensors is decreased by $\frac{5}{2}(p-1)$ to $\frac{1}{2}(5p-3)$.

Fig. \ref{crossg13} depicts an example of the $\mathbb{Z}^2$ cross array where $\mathbf{G}\mathbf{x}=(0, 6.5)^T$ and the number of sensors is decreased to 31. 
Likewise, in the $A_2$ case, it can be observed that after removing sensors below $x$ axis in the larger subarray and symmetrically extending the smaller subarray, 
%the points with $d'_y \geq \frac{1}{2}p$ are redundant since we can always find their corresponding symmetric points with respect to $(0, \frac{1}{2}p)$. 
the total number of sensors becomes $\frac{1}{3}(8p-5)$ which is $\frac{7}{3}(p-1)$ fewer than HSCRT. The reduction difference between $\mathbb{Z}^2$ and $A_2$ cross arrays is caused by the different shapes of the Voronoi cells when extending the smaller subarray, i.e., the Voronoi cell of the former is square whereas it is hexagonal for the latter. Fig. \ref{crosse13} depicts an example of $A_2$ cross array, from which it can be observed that the total number of sensors is reduced to 33.

Recall that the fragility is defined as the ratio of the number of essential sensors to the total number of sensors \cite{liu2018SAM}, i.e., 
$$\text{Fragility}=\frac{\text{The number of essential sensors} }{\text{Total number of sensors }}.$$
Thus it can be calculated that the fragile parameters are $0.26$ for hole-free $\mathbb{Z}^2$ and $0.30$ for hole-free $A_2$. By the two economic maximization methods, fragility surges to $1$ for all sliced CRT and cross arrays, i.e., all sensors are essential, which implies that the new arrays achieve \emph{maximally economic} while the HSCRT arrays are more robust. Table \ref{sum arrays} lists the sensor number and the fragility of the proposed array configurations for convenience.   

\begin{table}[tb]
\vspace{1em}
\caption{Summary of proposed sensor configurations}
\centering
\begin{tabular}{|l||*{2}{c|}}\hline
% \backslashbox{Array}
&\makebox[6.5em]{Senor Number}
&\makebox[6.5em]{Fragility}
\\\hline\hline
Hole-free $\mathbb{Z}^2$\cite{li2018part1} &  $5p-4 $ &  0.26\\\hline
Hole-free $A_2 $\cite{li2018part1}  & $5p-4 $ & 0.30  \\\hline
T array &  $ 3p-2 $ &  1\\\hline
Spinner array & $ 3p -2 $  & 1   \\\hline
$\mathbb{Z}^2$ cross  &  $1/2(5p-3)$ &  1   \\\hline
$A_2 $ cross & $1/3(8p-5)$ & 1  \\\hline
\end{tabular}
\label{table}
\vspace{1em}
\label{sum arrays}
\end{table}

\section{Applications to 2D DOA Estimation}\label{Applications to DOA Estimation}

This section verifies the feasibility of the proposed arrays including T, spinner, $\mathbb{Z}^2$ cross and $A_2$ cross in both passive sensing and active sensing scenarios corresponding to difference and sum coarrays respectively. A transformation technique will be exploited for converting the coarray from an arbitrary algebraic lattice to a uniformly distributed array, to which spatial smoothing and subspace-based methods can be applied.

\subsection{Passive Sensing}
In the passive sensing scenario, the data model is similar to that in \cite{pal2010nested, pal2012nested, liu2017hourglass}. Suppose $K$ uncorrelated narrow-band sources $s_1(t), s_2(t), \cdots s_K(t)$ impinge on a planar array whose $N$ antennas are placed on a 2D lattice $\mathcal{Z}$ with $\mathbf{z}=[z_{x}, z_{y}]^T \in \mathcal{Z}$ representing sensor positions.  
The received signals $x_1(t), x_2(t), \cdots, x_K(t)$ are dependent on the elevation angle set  $\mathcal{\theta}=\{ \theta_1, \theta_2, \cdots, \theta_K \}$ and azimuth angle set $\mathcal{\phi}=\{ \phi_1, \phi_2, \cdots, \phi_K \}$ and are corrupted by additive spatial wide-sense stationary (WSS) noise $\mathbf{n}(t) \sim \mathcal{N}(0, \eta^2)$. Therefore, the output of the antenna array can be expressed as:
\begin{equation}\label{xt}
\mathbf{x}(t) = \mathbf{A}\mathbf{s}(t)+\mathbf{n}(t).
\end{equation}
Here $\mathbf{x}(t), \mathbf{A}, \mathbf{s}(t)$ and $\mathbf{n}(t)$ represent the output vector, the steering matrix, the source vector and the noise signal vector respectively.
$\mathbf{A} \in \mathbb{C}^{N \times K}$ whose element at $q$th row and $k$th column is 
\begin{equation}\label{Aik}
[\mathbf{A}]_{q,k}=\exp\bigg(-i\frac{2\pi d}{\lambda}\mathbf{v}_{\theta,\phi}^{(k)}\mathbf{z}\bigg),
\end{equation}
where $\mathbf{v}_{\theta,\phi}^{(k)} = \sin\phi_k\big[\cos\theta_k, \, \sin\theta_k\big]$ and $d$ is the minimum distance of the interelement spacing. Since the noise $\mathbf{n}(t)$ is WSS and the transmitting signals are uncorrelated by assumption, the autocorrelation matrix of received data can be expressed as
\begin{equation}\label{Rx2}
\mathbf{R}_\mathbf{x}=E[\mathbf{x}\mathbf{x}^H]=\mathbf{A}\mathbf{R_s}\mathbf{A}^H + \eta^2 \mathbf{I}. 
% \mathbf{R}_\mathbf{x}=E[\mathbf{x}\mathbf{x}^H]=\mathbf{A}\mathbf{R_s}\mathbf{A}^H+\mathbf{n}\mathbf{n}^H 
\end{equation}
After vectorizing $\mathbf{R}_\mathbf{x}$ and removing the rows corresponding to zero elements in $\mathbf{s}$, the manifold vector can be written by 
\begin{equation}\label{vx2}
\mathbf{x}_z=\mathbf{A}_1\mathbf{s}_1+\mathbf{n}_1,
\end{equation}
where $\mathbf{A}_1=\mathbf{A}^* \odot \mathbf{A}$ ($\odot$ is Khatri-Rao Product operation and $*$ denote the Hermitian transpose), $\mathbf{s}_1 = [\sigma^2_1, \sigma^2_2, \cdots \sigma^2_K]^T$. Note that $\mathbf{x}_z$ has the same form as Equation (\ref{xt}); Therefore we can consider it as a new system input with the $N^2$-by-$K$ steering vector $\mathbf{A}_1$, the source vector $\mathbf{s}_1$ and the noise vector $\mathbf{n}_1$. Substituting (\ref{Aik}) into (\ref{Rx2}) and performing the vectorization results in
\begin{equation}\label{A1nk}
[\mathbf{A}_1]_{j,k}=\exp \bigg(i\frac{2\pi d}{\lambda} \mathbf{v}_{\theta,\phi}^{(k)} \mathbf{d}_{m,n} \bigg)
\end{equation}
%$\mathbf{v}_{\theta,\phi}=\sin(\theta)\big[\cos(\phi) \quad \sin(\theta)\big]$
where $\mathbf{d}_{m,n}$ is the $(m,n)$th element in difference coarray set $\mathcal{D}$
% $$ \mathcal{D}=\{ \mathbf{d} :  \mathbf{d} = \mathbf{z}_m + \mathbf{z}_n \} $$
whose cardinality limits the maximum DOF. Note that in general $\mathbf{d}_{m,n}$ is not an integer vector since the sensor location set $\mathcal{Z}$ can have non-integer elements.

As a quantification of occurrences of $\mathbf{d}_{m,n}$, the weighting function $w(\mathbf{d})$ is defined as the number of identical pairs occurring in $\mathcal{D}$:
\begin{equation}\label{wd}
w(\mathbf{d})=
\left|\{
\begin{array}{c|c}
\mathbf{d} = \mathbf{z}_m - \mathbf{z}_n  & \mathbf{z}_m, \mathbf{z}_n \in \mathcal{Z}.
\end{array}
\right\}|,
\end{equation}

With the presence of mutual coupling, the antenna array output can be rewritten by incorporating a mutual coupling matrix $\mathbf{C}$ into (\ref{xt}), i.e., 
\begin{equation}\label{xtc}
\mathbf{x}(t) = \mathbf{C} \mathbf{A}\mathbf{s}(t)+\mathbf{n}(t).
\end{equation}
% In addition to  conventional
Considering assumptions underlying the circuit model of dipole antennas, we assume that the absolute values of mutual coupling coefficients are inversely proportional to Euclidean norms of difference vectors of the planar array \cite{liu2016super, svantesson1999direction, friedlander1991direction, king1957mutual}. Under this assumption, the design method with smaller values of $w(\mathbf{d})$ at close sensor separations (shorter length of $ \mathbf{d} $) performs better since the effect of mutual coupling is mitigated as the increasing of the interelement spacing. In other words, the sparser the array is, the lower electromagnetic interactions will be.

\subsection{MIMO radar}

Let us consider a MIMO radar consisting of $M$ transmitting antennas allocated at lattice points $\mathbf{z}_m \in \mathcal{Z}_I$ and $N$ receiving antennas allocated at $\mathbf{z}_n \in \mathcal{Z}_J$ whose radiation patterns are represented as $C_E(\theta,\phi)$ with $\theta$ and $\phi$ being elevation and azimuth  respectively. Thus the radiation pattern of an antenna array with identical elements can be given by
\begin{equation}\label{C}
    C(\theta,\phi)=C_E(\theta,\phi)AF'(\theta,\phi),
\end{equation}
where  $AF'(\theta,\phi)$ denotes the normalized array factor $AF(\theta,\phi)$ of this antenna element:
\begin{equation*}
 AF'(\theta,\phi)=|AF(\theta,\phi)/AF(\theta,\phi)_{\text{max}}|.
\end{equation*}
The MIMO radar system  possesses the two-way radiation pattern $C_{MN}$ which is the product of the transmitting radiation pattern and the receiving antenna pattern, i.e., 
\begin{equation}\label{CMN}
    C_{MN}=C_M(\theta,\phi)C_N(\theta,\phi),
\end{equation}
where the array factors of transmitting and receiving arrays are given by 
\begin{equation}\label{AFM}
  AF_M(\theta,\phi) = \sum_{m=1}^M\mathbf{I}_{m} \exp\Big(j\frac{2\pi d}{\lambda}\mathbf{z}_m\sin(\theta)\cos(\phi)\Big)
\end{equation}
and
\begin{equation}\label{AFN}
  AF_N(\theta,\phi) = \sum_{n=1}^N\mathbf{I}_{n} \exp\Big(j\frac{2\pi d}{\lambda}\mathbf{z}_n\sin(\theta)\sin(\phi)\Big)
\end{equation}
respectively, where $\mathbf{I}_{m}$ ($\mathbf{I}_{n}$) is determined by the weights of the $m$th transmitter (the $n$th receiver) \cite{dahl2015comparison,gilbert2007antenna,bliss2003multiple,harter20132}. 
From (\ref{C}) and (\ref{CMN}), the two-way radiation pattern can be rewritten using the normalized array factors of transmitters and receivers:
\begin{equation}
    C_{MN}(\theta,\phi)=C_E(\theta,\phi)^2AF'_M(\theta,\phi)AF'_N(\theta,\phi).
\end{equation}
Therefore we can consider a virtual array with the normalized array factor $AF'_{MN}(\theta,\phi)=AF'_M(\theta,\phi)AF'_N(\theta,\phi)$. 
% It can be readily verified that virtual array factor can be represented by $AF_{MN}(\theta,\phi)=AF_M(\theta,\phi)AF_N(\theta,\phi)$. 
Substituting (\ref{AFM}) and (\ref{AFN}) to $AF_{MN}(\theta,\phi)$ results
\begin{equation*}
  AF_{MN}(\theta,\phi)= \sum_{m=1}^M \sum_{n=1}^N \mathbf{I}_{m} \mathbf{I}_{n}\exp\Big(j\frac{2\pi d}{\lambda} \mathbf{v}_{\theta,\phi} \mathbf{s}_{m,n}\Big),
\end{equation*}
where $\mathbf{v}_{\theta,\phi}=\sin(\theta)\big[\cos(\phi) \quad \sin(\theta)\big]$ and $\mathbf{s}_{m,n}$ is the $(m,n)$th element in the sum coarray set, which can be defined as follows: 
\begin{deft} [Sum coarrays of CRT arrays]
The sum virtual coarray $\mathcal{S}$ generated by a CRT array is given by:
\begin{equation*}\label{D}
\mathcal{S} = \{ \mathbf{z}_m + \mathbf{z}_n \mid  \mathbf{z}_m \in \mathcal{Z}_I ,  \mathbf{z}_n \in \mathcal{Z}_J \},
\end{equation*}
\end{deft}
where $\mathcal{Z}_I$ and $\mathcal{Z}_J$ are subarrays based on $\sigma(\mathcal{I})$ and $\sigma(\mathcal{J})$ respectively. For example, for cross arrays, $\mathcal{Z}_I = \mathcal{Z}_1 \cup \{  (z_x, z_y)  \, | \, z_x=2x_x-h_x, z_y=2x_y-h_y, (h_x, h_y)\in \mathcal{H}\}  $  and $\mathcal{Z}_J=\{ \mathbf{z}= (z_x, z_y) \; | \; \mathbf{z} \in \mathcal{Z}_2,  0 < z_y < x_x \}$.
The cardinality of $\mathcal{S}$ limits the maximum number of targets that can be detected. Note that when they are applied to MIMO radar, the HSCRT arrays (\cite[Definition 6]{li2018part1}) inherit all the properties derived in the passive sensing case including the hole-free property because their sum coarrays are identical to difference coarrays despite the fact that one subarray acts like a transmitter and the other is employed as a receiver.

\subsection{ Hexagon-to-rectangular Transformation in 2D Spatial Smoothing }
 
The conventional angle estimation techniques such as MUSIC and ESPRIT require uniformly distributed coarrays  \cite{schmidt1986multiple,roy1989esprit,rao1989performance,zoltowski1996closed}. However, both sum and difference coarrays of CRT arrays are algebraic lattices which are not uniform rectangular arrays in most cases.
Therefore, it is adequate to utilize the hexagon-to-rectangular transformation introduced in \cite{tian1998doa} that extends the direct-solution DOA techniques to other array geometries such as $A_2$ arrays. 

\begin{figure}[tb]
  \begin{center}
      \subfigure[]{\includegraphics[height=2.1in]{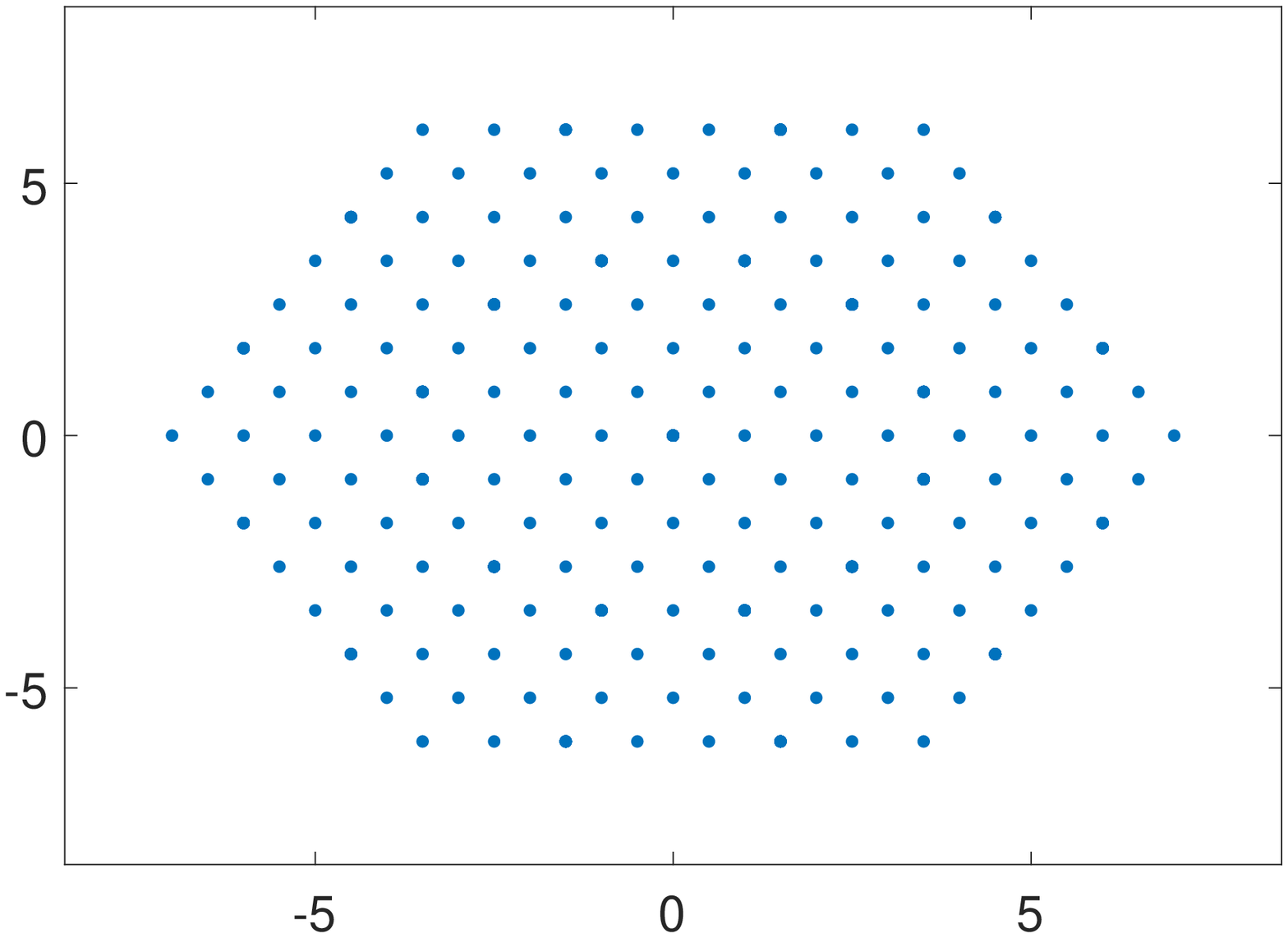}}
      \subfigure[]{\includegraphics[height=2.1in]{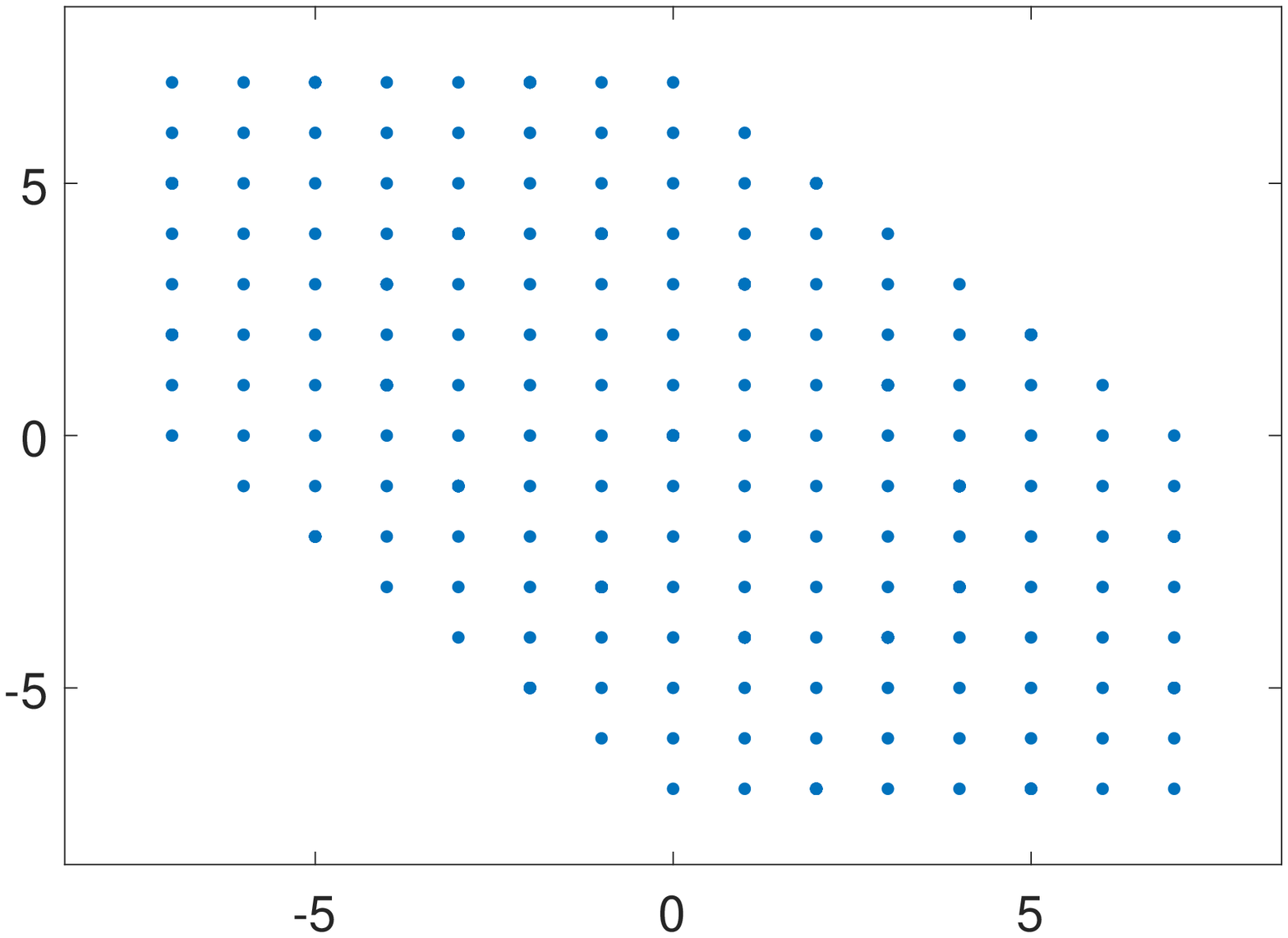}}
    \caption{An example of an $A_2$ array (a) and its equivalent array on $\mathbf{u}$ space (b).}
    \label{rotation3}
  \end{center}
\vspace{-6mm}
\end{figure}

Let $\mathcal{D}_{C,E}$ denote the continuous hexagonal part of the coarray and $l_R$ denote the circumradius of $\mathcal{D}_{C,E}$. Thus (\ref{A1nk}) can be rewritten as the following:
\begin{equation}\label{rotationAik}
[\mathbf{A}_1]_{d',k}=\exp \bigg(i\frac{2\pi d}{\lambda} \mathbf{v}_{\theta,\phi}^{(k)} \mathbf{G}[d'_x, d'_y]^T \bigg),
\end{equation}
where $\mathbf{G}$ is the generator matrix of the algebraic lattice $\Lambda = \sigma (\mathbb{Z}[q])$ of the ring $\mathbb{Z}[q]$, $d'_x, d'_y$ are rational integers, $-l_R\leq d'_x\leq l_R$ and  $-l_R\leq d'_y\leq l_R$. 
Thus the difference vector $\mathbf{d}$ can be expressed as 
$\mathbf{G}[d'_x, d'_y]^T$ for all $\mathbf{d} \in \mathcal{D}_{C,E}$.
The transformation from an arbitrary lattice to an equivalent rectangular array is realized by introducing an incident wave vector in the $\mathbf{u}$ space. For every source $k$, its wave vector is defined by: 
%$\frac{2\pi}{\lambda}\mathbf{u}_{\theta, \phi}^{(k)}$ where $\mathbf{u}_{\theta, \phi}^{(k)}$ is expressed as 
\begin{equation}\label{uk}
\mathbf{u}_{\theta, \phi}^{(k)}=\sin\phi_k[\cos\theta_k, \, \sin\theta_k]\mathbf{G}.
\end{equation}
By substituting (\ref{uk}) to (\ref{rotationAik}), the steering vector of the rectangular coarray in the $\mathbf{u}$ space can be written as 
\begin{equation*}
[\mathbf{A}_1]_{d',k}=\exp \bigg(-i\frac{2\pi d}{\lambda}\mathbf{u}_{\theta, \phi}^{(k)}[d'_x, d'_y]^T \bigg).
\end{equation*}
Here the affine transformation for an arbitrary point $\mathbf{d}=(d_x, d_y)$ with Cartesian coordinate system to the $\mathbf{u}$ space is 
\begin{equation*}
    \mathbf{d}'=\mathbf{G}^{-1}\mathbf{d}.
\end{equation*}
For example, the transformation from a $A_2$ lattice to a uniformly distributed array is realized by: 
\begin{equation}\label{u}
  \begin{pmatrix}
    d_x'  \\   d'_y
  \end{pmatrix}=
\begin{pmatrix}
    1  & -\frac{1}{\sqrt{3}} \\ 0 & \frac{2}{\sqrt{3}} 
  \end{pmatrix}
\begin{pmatrix}
    d_x  \\   d_y
  \end{pmatrix},
\end{equation}
as illustrated in Fig. \ref{rotation3} where $\mathcal{D}_{C,E}=\Lambda \cap \mathcal{V}(14A_2)$.
In this case, after the affine transformation, $\mathcal{D}_{C,E}$ becomes a parallelogram-like polygon in the $\mathbf{u}$ space whose edges are $(-l_R, 0)$, $(-l_R, l_R)$, $(0, l_R)$, $(l_R, 0)$, $(l_R, -l_R)$, $(0,-l_R)$. 
Note that if the subarrays are generated by coprime integer matrices, according to CRT, the generator matrix of their coarray is an identity matrix, implying that this coarray is a uniformly distributed array. Thus all $\mathbb{Z}^2$-based arrays do not require this transformation.

In short, we transform an arbitrary difference coarray described by an algebraic lattice in the Cartesian system with wave vector $\mathbf{v}_{\theta, \phi}^{(k)}$ to an equivalent parallelogram-like array based in the $\mathbf{u}$ space whose wave vector is $\mathbf{u}_{\theta, \phi}^{(k)}$. Considering 
%the passive sensing scenario where the
the estimation process formulated by the virtual array model (\ref{vx2}) with a larger array aperture and enhanced DOF, the rank of the coarray output covariance matrix $\mathbf{R}_\mathbf{z}=\mathbf{x}_z\mathbf{x}_z^H$ is one because of the vectorization on $\mathbf{R}_{\mathbf{s}}$, i.e., all receiving sources of virtual array become coherent. As a result, it is infeasible to apply subspace-based methods.
Herein, spatial smoothing is done using parallelogram-like polygon arrays transformed by the aforementioned technique, after which the estimated DOAs are relocated back to the Euclidean space using Cartesian coordinates. 
%Fig. \ref{rotation3} offers an illustration of the affine transformation on the difference coarray of $A_2$ sensor array. 

% Analogously to $\mathbb{Z}^2$  arrays, 
Let us define the $j$th subarray in the $\mathbf{u}$ space by
\begin{equation}
\begin{aligned} \label{SiE1}
      &\mathcal{S}_{j}=\{ (d'_x, d'_y)  \quad | \quad j -l_R \leq d'_x \leq j, \\
      & -j \leq d'_y \leq l_R - j  \},\\
\end{aligned}
\end{equation}
where $0 \leq j \leq l_R $. Let $\mathbf{v}_j$ denote the array manifold vector of the $j$th subarray. By the spatial smoothing, the covariance matrix can be calculated as
%Using the aforementioned spatial smoothing technique demonstrated in the previous subsection, the covariance matrix can be calculated by
\begin{equation}\label{Rsmooth}
\mathbf{R}_{\text{smooth}}=\frac{1}{l_R+1}\sum_{j=0}^{l_R}\mathbf{v}_{j}\mathbf{v}_{j}^H,
\end{equation}
whereby direct-solution estimation schemes like Unitary-ESPRIT can be performed. However, it can be observed from (\ref{Rsmooth}) that the number of spatial smoothing operations is only $l_R+1$, which 
leads to low estimation accuracy with small apertures of contiguous coarrays. To overcome this problem, we propose the following two data preprocessing methods. %the signal correlation

\subsubsection{Generalized Spatial Smoothing Method I}
\begin{figure}[tb]
  \begin{center}
      \subfigure[]{\includegraphics[height=2.2in]{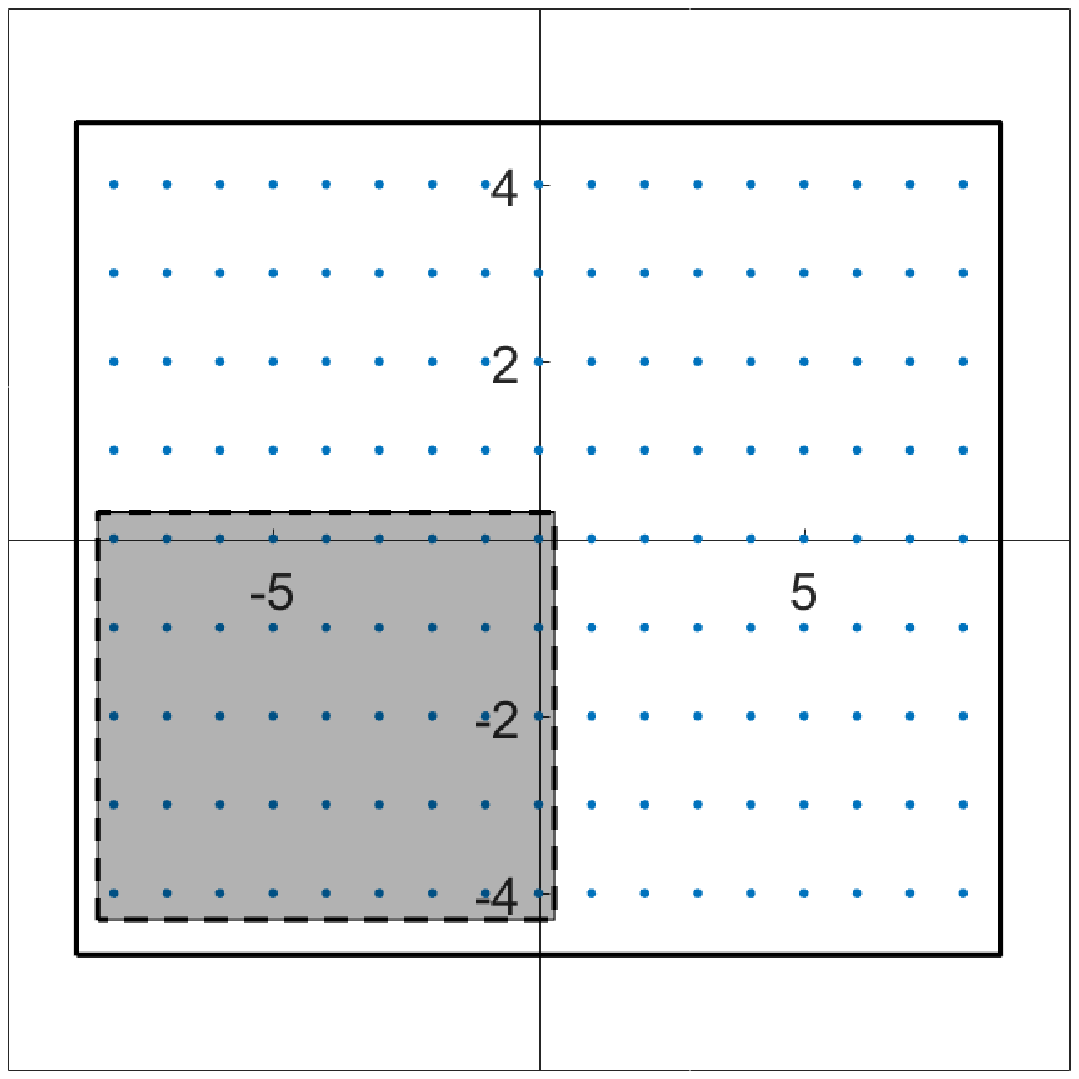}\label{smallsquared}}
      \subfigure[]{\includegraphics[height=2.2in]{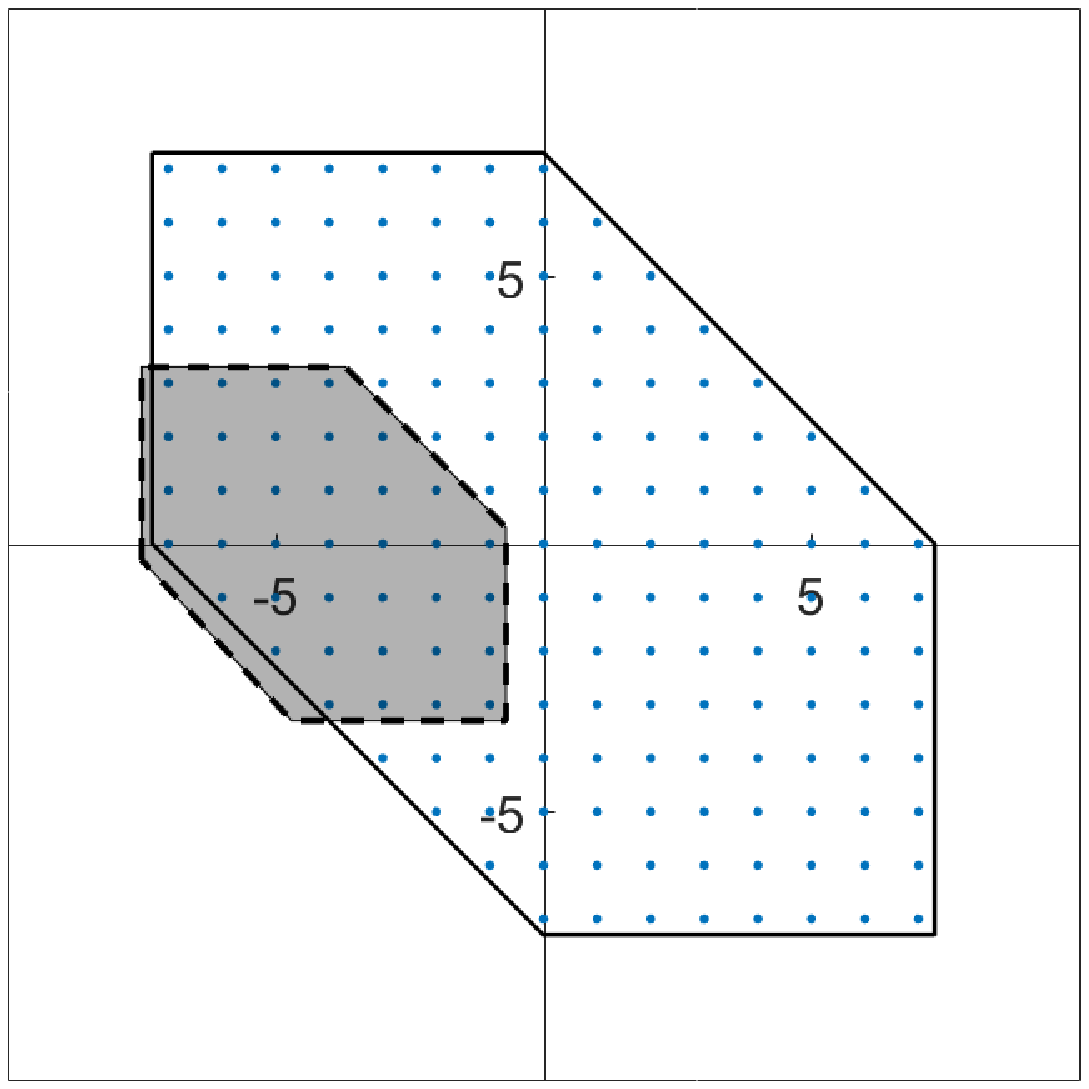}\label{smallhexagonald}}
    \caption{Spatial smoothing methods in the $\mathbf{u}$ space for $p=13$ : rectangular coarray (a) and parallelogram-like coarray (b) with the first subarray within gray shade.
    %the contiguous difference coarray aperture is shown in black cell with virtual elements in blue dot inside.
    }
    \vspace{-2em}
    \label{small}
  \end{center}
\end{figure}

In the equivalent $\mathbf{u}$ space, we select the rectangular contiguous part of the transformed virtual coarray with length $2x_g$ and width $2y_g$ as depicted in Fig. \ref{smallsquared}. 
% which contains $M$ distinct elements as depicted in Fig. \ref{smallsquared}. 

The $(0,0)$th subarray is a $l_x$-by-$l_y$ rectangular within the whole continuous array aperture as shown in gray shade at the bottom left in Fig. \ref{smallsquared}. The $(i_1, i_2)$th subarray is constructed by extracting elements from the difference coarray $\mathcal{D} $ \cite{shan1985spatial,vaidyanathan2011theory,pal2010nested}:
\begin{equation} \label{SiG}
\begin{aligned} 
      &\mathcal{S}_{i_1,i_2}=\{ \mathbf{d}' = (d'_x, d'_y)^T \quad |  \\
      & -x_g + i_1 \leq d'_x \leq - x_g +l_x +i_1, \\
      & -y_g + i_2 \leq d'_y \leq -y_g +l_y +i_2  \}\\
\end{aligned}
\end{equation}
where $ 0 \leq i_1 \leq 2x_g-l_x$ and $ 0 \leq i_2 \leq 2y_g-l_y $. Let $I_x=2x_g-l_x+1$ and $I_y=2y_g-l_y+1$. The number of subarrays is $I_xI_y$ and the DOF of each subarray is $(l_x+1)(l_y+1)$. Next step is evaluating the output of the $(i_1, i_2)$th subarray by searching all differences in $ \mathcal{S}_{i_1,i_2}$ from $\mathcal{D}$ with the same $\mathbf{d}$ and then reprocessing the output data accordingly. Let $\mathbf{x}_{i_1,i_2}$ extract the matched elements in $\mathbf{x}_z$ from (\ref{vx2}) to form a new output vector:
\begin{equation}
\mathbf{x}_{i_1,i_2} = \mathbf{A}_{i_1,i_2}\mathbf{s}_1 + \mathbf{n}_{i_1,i_2}.
\end{equation}
In practice, if there is more than one element in $\mathbf{x}_z$ that correspond to the same difference $\mathbf{d}$, we take the average over these elements, i.e.,
\begin{equation}
\mathbf{x}_{i_1,i_2}(\mathbf{d}) = \frac{1}{w(\mathbf{d})}\sum_{j=1}^{j=w(\mathbf{d})} \mathbf{x}_{z,j}(\mathbf{d}),
\end{equation}
where $1 \leq j \leq w(\mathbf{d}) $ and $\mathbf{x}_{z,j}(\mathbf{d}) \in \mathbf{x}_z$ denotes the rows of $\mathbf{x}_{z}$ that correspond to the difference vector $\mathbf{d}$ and there are a number $w(\mathbf{d})$ of such rows.  
Note that $\mathbf{n}_{i_1,i_2}$ is a vector with all zeros entries except $\eta^2$ at the entries corresponding to
$\mathbf{0}$ element in the difference coarray, i.e., $\mathbf{z}_m-\mathbf{z}_n = \mathbf{0}$; $\mathbf{A}_{i_1,i_2}$ is a $ (l_x+1)(l_y+1)$-by-$K$ matrix defined by
\begin{equation}\label{Ai1i2}
\mathbf{A}_{i_1,i_2} = [\mathbf{A}_x \mathbf{D}_1( \mathbf{A}_y ), \mathbf{A}_x \mathbf{D}_2 (\mathbf{A}_y), \cdots ,\mathbf{A}_x \mathbf{D}_{l_y} (\mathbf{A}_y)]^T,
\end{equation}
Here $ \mathbf{D}_{r} (\mathbf{A}_y)$ is a diagonal matrix of dimension $l_y$ whose diagonal entries are the $r$th row of $\mathbf{A}_y$. $\mathbf{A}_x$ and $\mathbf{A}_y$ are the steering matrices of the $(l_x+1)$ sensors on $x$ axis and of the $(l_y+1)$ sensors on $y$ axis respectively, which are expressed as two Vandermonde matrices, namely
\begin{equation}
[\mathbf{A}_x]_{a_1,k}=\exp \Big(-i\frac{2\pi d}{\lambda}\sin\phi_k
\sin\theta_k (-x_g+a_1+i_1 \Big), 
\end{equation}
\begin{equation}
[\mathbf{A}_y]_{a_2,k}=\exp \Big(-i\frac{2\pi d}{\lambda}\sin\phi_k
\cos\theta_k (-y_g+a_2+i_2\Big),
\end{equation}
where $ 0 \leq a_1 \leq l_x $ and $ 0 \leq a_2 \leq l_y$. Therefore, spatial smoothing can be performed by averaging the output data covariance of all virtual rectangular subarrays, i.e., 
\begin{equation}
\mathbf{R}_{\text{smooth}}=\frac{1}{I_xI_y}\sum_{i_1=0}^{I_x-1}\sum_{i_2=0}^{I_y-1}\mathbf{x}_{i_1,i_2}\mathbf{x}_{i_1,i_2}^H
\end{equation}
Even if $\mathbf{A}_x$ and $\mathbf{A}_y$ are full-rank Vandermonde matrices, it can be observed from (\ref{Ai1i2}) that the steering vector $\mathbf{A}_{i_1,i_2}$ may be rank deficient which leads to identification issues when the subspace-based algorithms are applied using $\mathbf{R}_{\text{smooth}}$. \cite{pal2012nested} provides a sufficient condition on the unique identifiability, namely the objective sources must be fewer than or equal to $(l_x+1)( l_y+1)$ distinct pairs.

\subsubsection{Generalized Spatial Smoothing Method II}
\begin{figure}[tb]
      \centering
     %\framebox{\parbox{3in}{content}}
     \includegraphics[scale=0.4]{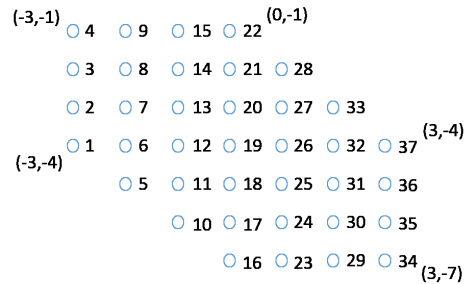}
    %\includegraphics[scale=0.4]{Figs/Jx1selection.png}
    % \vspace{-2mm}
      \caption{A graphical representation of elements in $\mathcal{S}_{0,0}$ with $l_R=7$ and $l_p=3$ from which $\mathcal{J}_{x1}$ can be selected by matrix $\mathbf{J}_{x1}$. Physical positions of sensors on the edges are shown by coordinates in the $\mathbf{u}$ space.}
      \label{Jx1selection}
      \vspace{-3mm}
  \end{figure}

Likewise, the parallelogram-like subarray structure is shown in Fig. \ref{smallhexagonald} where each subarray is shifted along $x$ axis or $y$ axis by a unit till it reaches the boundary. Precisely, let $l_p \in \mathbb{Z}$ denote the aperture of parallelogram-like subarrays and $0 < l_p < l_R$. For instance, the $(0,0)$th subarray is bounded by $(-l_R, 0)$ $(-l_R,l_p)$, $(l_p-l_R, l_p)$, $(2l_p-l_R, 0)$, $(2l_p-l_R, -l_p)$ and $(l_p-l_R, -l_p)$. Spatial smoothing is done by averaging these subarrays in the $\mathbf{u}$ space of which $(i_1,i_2)$th subarray is defined as:
%$(0,-l_R)$, $(l_x, -l_y)$, $(l_x, l_y-l_R)$, $(0, 2l_y-l_R)$, $(-l_x, 2l_y-l_R)$ and $(-l_x, l_y-l_R)$.
\begin{equation} \label{SiE2}
\begin{aligned} 
      \mathcal{S}_{i_1,i_2}=
      &\{ \mathbf{d}' = (d'_x, d'_y) \quad | \quad 
      \mathbf{d}' \in \mathcal{D}'_{C,E}  \\
      & d'_x + d'_y +l_R -i_1-i_2\geq 0, \\
      & -l_R+i_1 \leq d'_x \leq 2l_p - l_R + i_1, \\
      & l_p+i_2 \leq d'_y \leq -l_p + i_2, \\
      & d'_x + d'_y +l_R-2l_p -i_1-i_2 \leq 0  \},  \\
\end{aligned}
\end{equation} 
where $\mathcal{D}'_{C,E} = \{ \mathbf{d}'=\mathbf{G}^{-1}\mathbf{d} \; | \; \mathbf{d} \in \mathcal{D}_{C,E} \}$, $ 0 \leq i_1 \leq 2l_R$ and $-l_R \leq i_2 \leq l_R$. The DOF is obtained by calculating the number of elements in $\mathcal{S}_{i_1,i_2}$, i.e., $\text{DOF} = |\mathcal{S}_{i_1,i_2}| = 3l_p^2 + 3l_p + 1$. Note that direct searching algorithms such as MUSIC can still apply on the parallelogram-like subarrays. However, the ESPRIT-like algorithms which are computationally easier require URA array geometries since their array manifold formulations are based on the extension of ULAs. Induced by the translational invariance structure of virtual coarray, the estimation parameters can be calculated by exploiting the underlying shift invariance property after spatial smoothing.
%Inspired by
Similar to the vector selection technique employed on physical sensors \cite{van2004optimum,tian1998doa,roy1989esprit}, we define the array selection matrix $\mathbf{J}_{x1}$ as a $R$-by-$Q$ matrix whose elements in the $r$th row is defined by
\begin{align}
  [\mathbf{J}_{x1}]_{r,q} = \begin{cases}
      1 & \text{if $q \in \mathcal{J}_{x1}$} \\
      0 & \text{otherwise}
    \end{cases},
\end{align}
where $\mathcal{J}_{x1}$ is an $R$ element integer set composed of the numerical orders of elements within the selected hexagon. For example, the $(0, 0)$th subarray in the case of $p=13$ is shown in Fig. \ref{Jx1selection} and the elements in $\mathcal{J}_{x1}$ are $\mathcal{S}_{0,0} \backslash \{ 22, 28, 33, 34, 35, 36, 37 \}$, where $\backslash$ denotes the set-theoretic difference operation. 

Likewise, $\mathcal{J}_{x2}$ is defined by shifting $\mathcal{J}_{x1}$ along the positive direction of $x$ axis. $\mathcal{J}_{y1}$ and $\mathcal{J}_{y2}$ are defined by shifting one unit along negative and positive directions of $y$ axis respectively. It can be verified that these selected subarrays are symmetric and contain the estimation information. Therefore, unitary ESPRIT can be applied after which estimations of wave vectors in the Cartesian system can be achieved by left-multiplying the transpose vector, namely $\mathbf{G}$. 
Both generalized spatial smoothing methods are feasible and significantly increase the number of sources that can be detected. Considering variations of coarray apertures obtained by different ideals, we shall choose the method with a larger continuous region in order to reprocess more data.

\section{Numerical Results}\label{Numerical Results}

In this session, the feasibility of the proposed arrays including T, spinner, $\mathbb{Z}^2$ cross and $A_2$ cross will be explicitly demonstrated in the context of the passive sensing and MIMO radar, where the generalized spatial smoothing I and II will be employed for $\mathbb{Z}^2$-based and $A_2$-based arrays respectively to exploit more elements in virtual coarrays. For illustration purposes, all proposed arrays are generated from $p=13$ with notations shown in Fig. \ref{sector} and Fig. \ref{cross}, and compared with 2D nested arrays \cite{pal2012nested} with $N_1=N_2=3$. The numbers of sensors are only 31 for $\mathbb{Z}^2$ cross and 33 for $A_2$ cross while all the other arrays require a number 37 of sensors. It is noteworthy that the perimeters and the areas of all $A_2$-based arrays ($A_2$ cross and spinner) are $86\%$ of those $\mathbb{Z}^2$-based arrays ($\mathbb{Z}^2$ cross and T) \cite{li2018part1}.  
The minimum interelement distance $d$ is chosen to be $1/2\lambda$.

\subsection{Direction of Arrival Estimation}
The numerical results are presented to compare the DOA estimation performances among the proposed arrays, where all antennas are employed as receivers.

$K=6$ narrow-band uncorrelated sources modeled as Gaussian are buried in temporally and spatially WSS noise ($\text{SNR}=0$dB). Impingement angles are randomly distributed on $[-\pi, \pi]^2$. The input data is formulated as in (\ref{xtc}) with maximum mutual coupling coefficient being $0.2$, and each antenna can be affected by adjacent antennas within a circular aperture of length $3d$. 
Generalized spatial smoothing method I and II are applied by choosing $x_g=y_g=l_x=l_y=7$ for the former and $l_R=7$ and $l_p=3$ for the latter. The root-mean-square error (RMSE) is calculated by 
% pairwisely obtaining the errors between real and estimated DOAs then 
averaging the errors over all sources and over all independent simulations, i.e., %$49$ times ($S=49$)
\begin{equation}
\text{RMSE}=\sqrt{\frac{1}{KT}\sum^T_{j=1}\sum^K_{k=1}\big((\bar{\theta}_k-\hat{\theta}_{k,j})^2+(\bar{\phi}_k-\hat{\phi}_{k,j})^2\big)},
\end{equation}
where $(\bar{\theta}_k, \bar{\phi}_k)$ and $(\hat{\theta}_{k,j}, \hat{\phi}_{k,j})$ are real and estimated DOAs respectively. $T=1000$ Monte Carlo trials are performed on all the proposed arrays.
 \begin{figure}[tb]
  \begin{center}
      \subfigure[]{\includegraphics[height=1.55in]{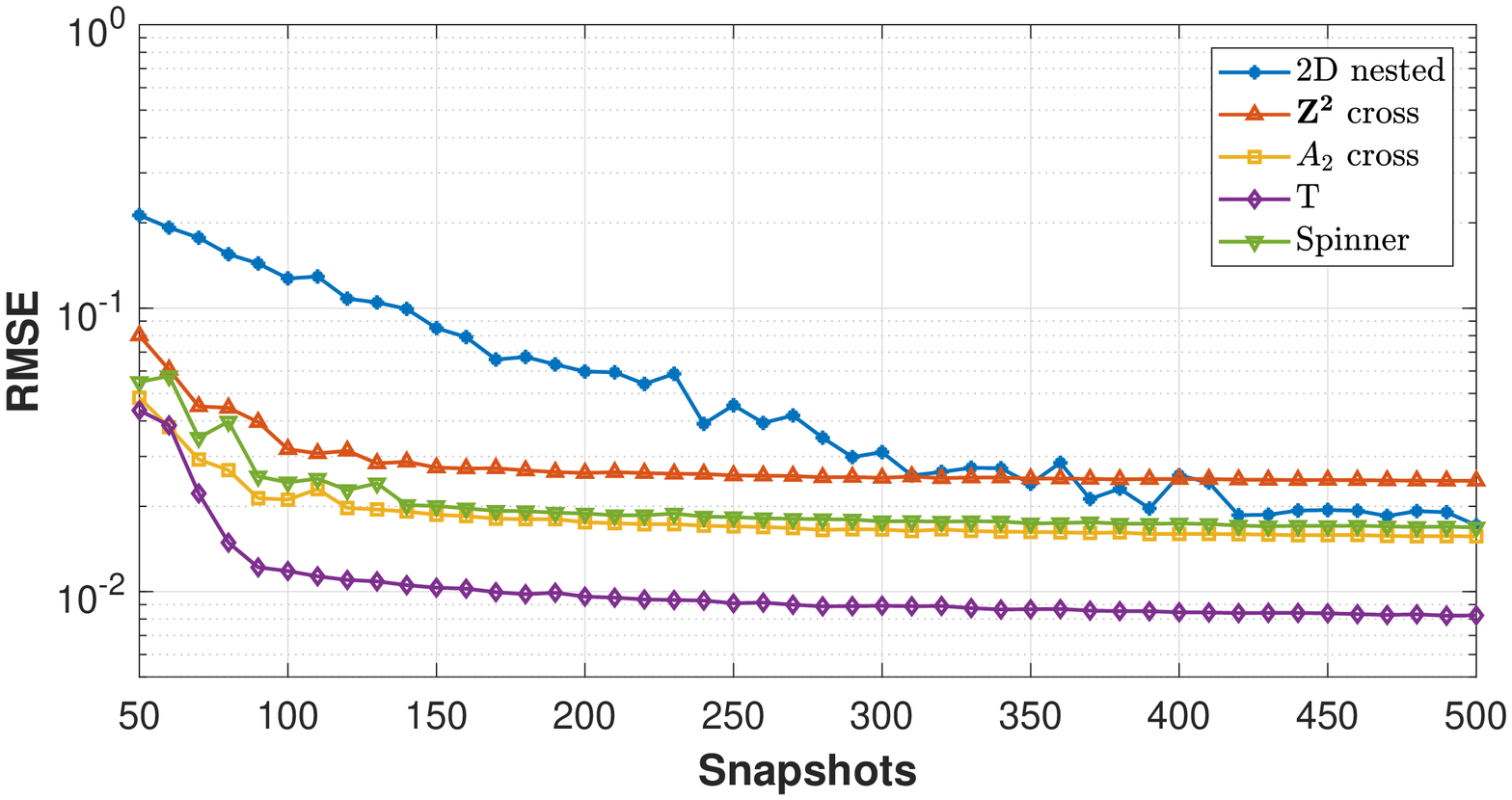}\label{snapshot_without_ura}}
      \subfigure[]{\includegraphics[height=1.55in]{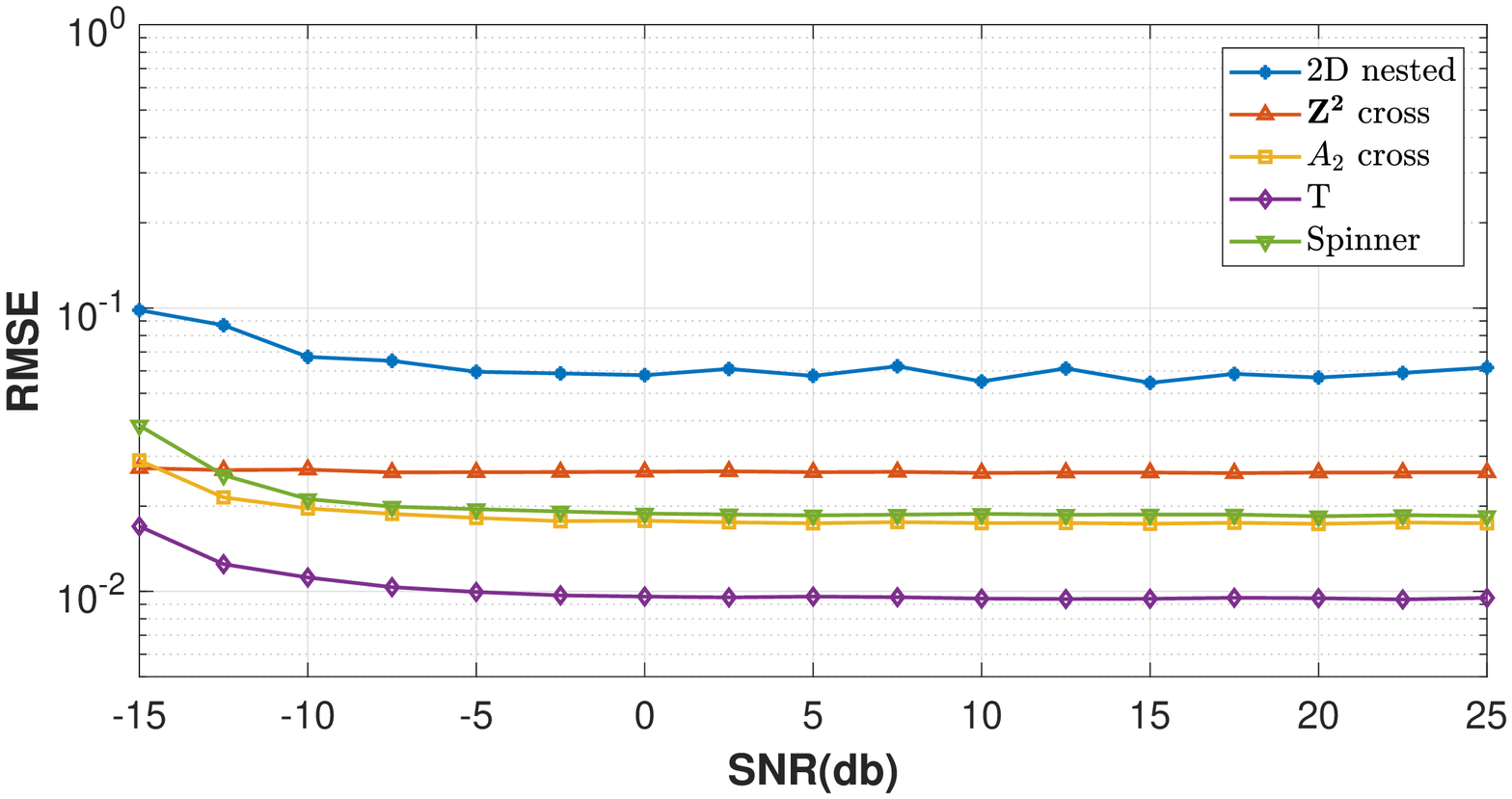}\label{snr_without_ura}}
    \vspace{-1mm}
    \caption{RMSE versus snapshots where SNR$= 0$dB (a) and RMSE versus SNR where $L=200$ (b). }
    \label{e}
  \end{center}
\vspace{-3mm}
\end{figure}

Fig. \ref{snapshot_without_ura} plots the RMSE values depending on the number of snapshots where the SNR is set to be $0$ dB. It is noticeable that all CRT arrays can manage DOA estimation within relatively small errors using only $L=50$ snapshots. With an increased number of snapshots, CRT-based arrays improve the estimation performance significantly. T and $A_2$ cross achieve superior performances when $L < 100$ whereas the difference between the spinner array and $A_2$ cross is less prominent for $L>100$. 
Fig. \ref{snr_without_ura} demonstrates the performances of CRT-based arrays by comparing the RMSE as a function of SNR. The minimum RMSE is achieved by T array, followed by the spinner array and $A_2$ cross.  
It can be observed that the proposed design methods have remarkably enhanced the estimation performance with fewer sensors.

\subsection{Radiation Pattern of MIMO Radar}
 \begin{figure}[tb]
  \begin{center}
    \includegraphics[height=1.85in]{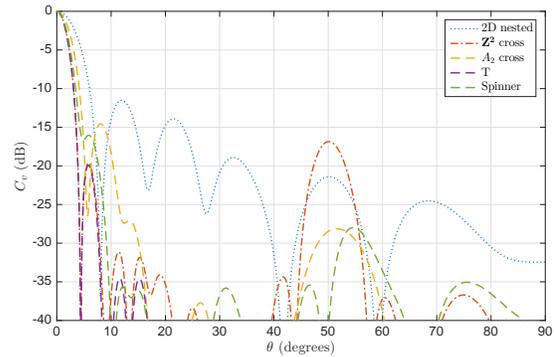}\label{rp32}
    \caption{Radiation patterns of T, spinner, $\mathbb{Z}^2$ cross, $A_2$ cross and the 2D nest array, where the azimuthal angle $\phi$ is set for the largest occurring side lobe.}
    \label{rp}
  \end{center}
\vspace{-6mm}
\end{figure}

Within the MIMO framework, the superiority of the proposed arrays is demonstrated in the context of the radiation patterns where sensors placed on the subset of $\sigma(\mathfrak{p}_1)$ (or equivalently, on $\mathcal{Z}_{I}$) are employed as transmitters and the rest as receivers. For the 2D nested array given in \cite{pal2012nested}, the dense array configuration corresponding to $N_1=3$ acts like the transmitting array and the sparse array with $N_2=3$ is the receiving array.

A comparison of MIMO concepts is shown in Fig. \ref{rp}. It can be measured that the side lobe suppression (SLS) of T array is dramatically reduced to $\text{SLS}_{A_2}=20$dB which is better than $\text{SLS}_{\text{nested}}=11.41$dB of the 2D nested array. As a result, the T array can target sources at a better separation. The least half power beam width is exhibited by T and $\mathbb{Z}^2$ cross, followed by spinner, $A_2$ cross and the 2D nested array. T array and $\mathbb{Z}^2$ array improve the angular resolution by approximately $3.6^\circ$ compared to the nested array. In short, the proposed CRT-based arrays outperform the known sparse array in terms of SLS and the angular resolution.

\section{Conclusion}\label{Conclusion}

This paper has explicitly demonstrated how the derivations of CRT arrays including T, spinner, $\mathbb{Z}^2$ cross and $A_2$ cross arrays can be applied to both active and passive sensing. 
A general approach to coprime lattices has been proposed based on Bezout's identity and the closed property of quadratic fields, which provides the possibility of exploiting all quadratic integers in number theory. 
By the underlying symmetry of ideal lattices, 
the number of sensors was significantly reduced from HSCRT
while the difference coarray aperture remains intact. We incorporated the hexagon-to-rectangular technique with 2D spatial smoothing along with matrix selections of coarrays, which extends subspace-based estimation algorithms to non-uniformly distributed sparse arrays. 
The proposed design methods provide sparse array geometries and increased DOF and significantly alleviate the mutual coupling effect. The numerical analysis illustrates the performance of the CRT-based arrays with variations of SNR and snapshots.                  

Future work will address the application of the proposed methods in the case of coherent signals.
Another direction in the future research is towards the optimization of coarray apertures by maximizing norms of the ideals in quadratic fields.

\begin{appendices}

\section{Proof of Theorem \ref{coprime integers}}\label{proof coprime integers}

By the quadratic formula, we can express $q$ and $\hat{q}$ by the following \cite{Oggier10,marcus1977number, li2018part1}:
\begin{equation}\label{q hatq}
      q = -\frac{1}{2}B + \frac{1}{2}\sqrt{B^2-4C} , \quad  \hat{q} = -\frac{1}{2}B - \frac{1}{2}\sqrt{B^2-4C}.
\end{equation}
Here $q$ and $\hat{q}$ can be real or complex numbers corresponding to real and imaginary quadratic field respectively.

According to Bezout's identity \cite{adhikari2003groups}, with $\{ 1, q \}$ being the basis, $m, n \in \mathbb{Z}[q]$ are coprime if and only if there exist $\alpha=\alpha_1+\alpha_2 q \in \mathbb{Z}[q]$ and $\beta = \beta_1 + \beta_2 q \in \mathbb{Z}[q] $ such that $\alpha m + \beta n = 1$. Because number fields are closed under multiplication and addition, $ \alpha m + \beta n $ is also an element in $\mathbb{Z}[q]$, if all four quadratic integers are in $\mathbb{Z}[q]$. Taking the norms of both sides of Bezout's equation results in: 
\begin{equation}\label{1=}
\begin{aligned}
    & 1=(\alpha m + \beta n)(\hat{\alpha} \hat{m} + \hat{\beta} \hat{n}) \\
    & = \text{N}(\alpha)\text{N}(m)+\text{N}(\beta)\text{N}(n)+\hat{\alpha}\beta\hat{m}n+\alpha\hat{\beta}m\hat{n},
\end{aligned}
\end{equation}
where $\hat{m}=m_1+m_2\hat{q}$ is the algebraic conjugate of $m$ and same with $\hat{n}$, $\hat{\alpha}$ and $\hat{\beta}$.
Similarly, because $m, n, \hat{m}, \hat{n}$ are all in $\mathbb{Z}[q]$, $\hat{m}n$ and $m\hat{n}$ are also elements in $\mathbb{Z}[q]$ and can be expressed as $\hat{m}n = \delta_1 + \delta_2q$ and $m\hat{n} = \delta_1 + \delta_2 \hat{q} $ where $\delta_1 = m_1n_1 - Bm_2n_1+Cm_2n_2$ and $\delta_2 = m_1n_2-m_2n_1$. 
Note that $q$ and $\hat{q}$ are roots of $X^2+BX+C=0$, thus $q + \hat{q} = -B$ and $q\hat{q}=C$.
Likewise, $\hat{\alpha}\beta = \gamma_1 + \gamma_2q$ and $\alpha\hat{\beta} = \gamma_1 + \gamma_2 \hat{q}$. Therefore, (\ref{1=}) can be rewritten as
\begin{equation}
    \begin{aligned}
    & \text{N}(\alpha)\text{N}(m)+\text{N}(\beta)\text{N}(n) +\delta_1(2\gamma_1 - B\gamma_2) \\
    & +\delta_2(-B\gamma_1-B^2\gamma_2-2C\gamma_2)=1,
    %& +\delta_2(\gamma_1A+\gamma_2(A^2-2B))=1
    \end{aligned}
\end{equation}
which indicates that $\text{N}(m)$, $\text{N}(n)$, $ \delta_1$ and $ \delta_2$ are coprime from Bezout's identity, i.e.,
\begin{equation}\label{gcdNm}
    \text{GCD}(\text{N}(m), \text{N}(n), \delta_1, \delta_2)=1,
\end{equation}
Since $\text{N}(m)\text{N}(n)= m\hat{m} n\hat{n}=(m\hat{n})(\hat{m}n)=(\delta_1 + \delta_2q)(\delta_1 + \delta_2 \hat{q})=\delta_1^2 - B\delta_1\delta_2 + C\delta_2^2$, the coprimality of all four integer elements is the same as the coprimality of the first three elements: 
\begin{equation*}
    \begin{aligned}
     & \text{GCD}(\text{N}(m), \text{N}(n), \delta_1, \delta_2)
       =\text{GCD}(\text{N}(m), \text{N}(n), \delta_1^2, \delta_2) \\
     & =\text{GCD}(\text{N}(m), \text{N}(n), \delta_1^2 - \text{N}(m)\text{N}(n), \delta_2) \\
     & =\text{GCD}(\text{N}(m), \text{N}(n), \delta_2(B\delta_1-C\delta_2), \delta_2) \\
     &  =  \text{GCD}(\text{N}(m), \text{N}(n), \delta_2).
    \end{aligned}
\end{equation*}
Likewise, it can also be verified that (\ref{gcdNm}) holds if and only if (\ref{gcd(N(m), N(n), m_2n_1-m_1n_2)=1}) holds.
This implies that both coprime conditions are necessary.

Next, we show that both of the necessary conditions are also sufficient. In other words, we prove the coprimality of $m$ and $n$ by assuming any of the coprimality conditions is satisfied.  
To begin with, let us assume (\ref{gcd(N(m), N(n), m_1n_1+m_2n_2)=1}) holds. By adding and subtracting $m_2n_2q$ to the last term in (\ref{gcd(N(m), N(n), m_1n_1+m_2n_2)=1}), this condition can be rewritten as 
\begin{equation}\label{gcdmnm2n2}
\text{GCD}(\text{N}(m),\text{N}(n),(m_1n_2+m_2n_2q)-(m_2n_1+m_2n_2q))=1,
\end{equation}    
Recalling that $m=m_1+m_2q$ and $n=n_1+n_2q$, (\ref{gcdmnm2n2}) can be simplified to 
\begin{equation*}
    \text{GCD}(\text{N}(m),\text{N}(n),n_2m-m_2n)=1.
\end{equation*}
According to Bezout's identity, there exists quadratic integers $\zeta$, $\eta$ and $\mu$ in $\mathbb{Z}[q]$ such that 
\begin{equation}\label{NmNn=1}
    \zeta \text{N}(m) + \eta \text{N}(n) + \mu (n_2m-m_2n) = 1. 
\end{equation}
Substituting $\text{N}(m)=m\hat{m}$ and $\text{N}(n)=n\hat{n}$ to (\ref{NmNn=1}) results 
\begin{equation*}
    m(\zeta \hat{m} + \mu n_2 ) + n (\eta \hat{n} - \mu m_2 ) =1,
\end{equation*}
therefore, $m$ and $n$ are coprime if (\ref{gcd(N(m), N(n), m_1n_1+m_2n_2)=1}) holds. 
Similarly, (\ref{gcd(N(m), N(n), m_2n_1-m_1n_2)=1}) can be rewritten as 
\begin{equation*}
    \text{GCD}(\text{N}(m), \text{N}(n), (n_1-Bn_2)m-qm_2n)=1,
\end{equation*}
which implies that there exist $\zeta'$, $\eta'$ and $\mu'$ in $\mathbb{Z}[q]$ such that 
\begin{equation*}
    \zeta' \text{N}(m) + \eta' \text{N}(n) + \mu' ((n_1 - Bn_2)m - qm_2n) =1,
\end{equation*}
which can be simplified as
\begin{equation*}
    m(\zeta' \hat{m} + \mu' (n_1-Bn_2)) + n(\eta' \hat{n} - \mu' m_2q)=1.
\end{equation*}
Summarizing, the coprimality conditions are not only sufficient but also necessary. 

\end{appendices}

\bibliographystyle{ieeetr}
\bibliography{reference}

\end{document}